\documentclass[11pt,a4paper]{JHEP3}

\usepackage{bm}
\usepackage{graphicx}
\usepackage{epsfig}

\title{Cosmology and Fermion Confinement in a Scalar-Field-Generated
Domain Wall Brane in Five Dimensions}

\author{Tracy R. Slatyer and Raymond R. Volkas \\ 
School of Physics \\
Research Centre for High Energy Physics \\
The University of Melbourne \\
Victoria 3010 Australia \\ E-mail:
\email{r.volkas@physics.unimelb.edu.au \\ tslatyer@fas.harvard.edu} }

\abstract{We consider a brane generated by a scalar field domain wall configuration in 4+1 dimensions,
interpolating, in most cases, between two vacua of the
field. We study the cosmology of such a system in the cases where the
effective four-dimensional
brane metric is de Sitter or anti de Sitter, including a
discussion of the bulk coordinate singularities present in the de-Sitter
case. We demonstrate that a scalar field kink configuration can support a brane with
dS$_4$ cosmology, despite the presence of coordinate singularities in
the metric. We examine the trapping of fermion fields on the domain wall for
nontrivial brane cosmology.}

\begin{document}

\section{Introduction}

\label{introduction}

Models incorporating more than $3+1$ dimensions have been of great interest to the
high-energy physics community from the early 1980's onward, and have been
investigated in considerable depth over
the last decade (see, for instance, Refs. \cite{rubakov+shaposhnikov,
large extra dimensions, RS1, RS2}). It was initially thought that
any such extra dimensions would need to be compactified to a small radius in order for
their effect on four-dimensional gravity to be consistent with current
experiments. However, the work of
Randall and Sundrum (\cite{RS1, RS2}) showed that effective
four-dimensional gravity could be recovered even in the case of
non-compact extra dimensions. 

The original Randall-Sundrum proposals, and much of the subsequent
work, dealt with Standard Model fields confined a priori to a 3+1-dimensional subspace of the
higher-dimensional space, with gravity confined to
this ``brane'' by a warped five-dimensional metric (reviews are given
in \cite{general reviews, perez-lorenzana}). In these models the
brane is simply the boundary between two domains in the bulk
spacetime, with the energy-momentum localised on the brane
corresponding to a discontinuity in the derivatives of the metric
across the brane. The ensuing cosmological evolution of the brane-bulk
system has been studied extensively in the literature
(\cite{basicRScosmo, padilla}, or \cite{padilla, dynamic domain walls}
for discussions of cosmology induced by moving branes). Many modifications and extensions to the initial RS
scheme have also been proposed: see, for example
\cite{padilla, dynamic domain walls, RSmodified, alternative-braneworlds, kaloper, flanagan-et-al}.

It has been suggested that such a 3+1-dimensional brane could also be realised as a smooth domain wall
that dynamically confines the fields of the Standard Model
(\cite{rubakov+shaposhnikov, ringeval, dando+davidson+george+volkas+wali, scalar field
solutions, randjbar-daemi+shaposhnikov, kehagias+tamvakis, dewolfe+freedman+gubser+karch}). 
General aspects of smooth analogues of the Randall-Sundrum model
have been studied in \cite{csaki+erlich+hollowood+shirman,
bronnikov+meierovich}. In
particular, given a scalar field potential with two neighbouring
minima, a domain wall soliton may form where the scalar field
interpolates between two regions of spacetime corresponding
asymptotically to vacua
of the field. This approach is
appealing because it provides a dynamical origin for a brane model
with one infinite extra dimension.

A number of authors have investigated solutions to the Einstein
equations for a five-dimensional warped (nonfactorisable) metric
coupled to a scalar field (\cite{ringeval,
dando+davidson+george+volkas+wali, scalar field solutions,
kehagias+tamvakis, dewolfe+freedman+gubser+karch,  soda, afonso,
flanagan+tye+wasserman2, gremm, davidson+mannheim, wang}; see also \cite{farakos, bogdanos} for
non-minimally coupled scalar-gravity models). The trapping of
fermion fields on a subspace in models with extra dimensions has also been explored
(\cite{rubakov+shaposhnikov, ringeval, randjbar-daemi+shaposhnikov, kehagias+tamvakis, koley+kar1,
koley+kar2, jackiw+rebbi, bajc+gabadadze, csaki+hubisz+meade, melfo}). 
However, most (but not all) of these investigations have only considered the
case where the effective 4D metric is Minkowskian, so they do not
explore the cosmology of these models. The purpose of this paper is to study
aspects of the cosmology of domain-wall-style branes.

In Sec. \ref{The equations of motion for a warped metric coupled to a scalar} 
we write down the Einstein equations for a
warped 5D metric based on an arbitrary 4D metric, with the
energy-momentum tensor given by the bulk scalar field. We then discuss
the properties of solutions to the equations of motion, in the cases where
the effective 4D metric is Minkowski, dS$_4$ and AdS$_4$.

After reviewing some examples of scalar field domain walls with 4D
Minkowski slices (Sec. \ref{scalar field kink solutions for a
Minkowski brane}), we present criteria for a smooth 5D metric containing no curvature
singularities with dS$_4$ slices in Sec. \ref{properties of solutions
to the Einstein equations with dS$_4$ cosmology}. In
Sec. \ref{Partially analytic solution with dS$_4$ slices and no
curvature singularities} we write down one such metric
analytically, and plot the corresponding scalar field and potential
for a particular parameter set. In Sec. \ref{locally localised
gravity} we present new analytic solutions for the 5D warped
metric with AdS$_4$ and dS$_4$ slices on
the brane, and the associated background scalar field, motivated by
thin-brane solutions that have been employed to demonstrate
localisation of gravity (\cite{karch+randall}).

Finally, in Sec. \ref{fermion trapping in warped metrics} we consider the trapping of fermions in a spacetime described
by a 5D warped metric (with dS$_4$ or AdS$_4$ brane cosmology) and supported by a kink-like scalar field domain
wall. In Sec. \ref{confinement of fermion zero mode} we examine the behaviour of the chiral fermion zero modes in
these various solution systems.

\section{The equations of motion for a warped metric coupled to a scalar}

\label{The equations of motion for a warped metric coupled to a
scalar}

 Suppose $g^{(4)}_{\mu \nu}$ is some general 4D metric. Let us consider a 5D warped metric
$g_{M N}$ of the simple form,
\begin{equation} g_{M N} = \pmatrix{ e^{- 2 \sigma(y)} g^{(4)}_{\mu \nu} & 0  \cr
0 & 1 }. \label{general warped metric}
\end{equation}
(In general, we shall use capitalised Roman letters to indicate
five-dimensional indices, and Greek letters for
four-dimensional indices.)
We use the sign conventions of \cite{weinberg}: in particular, the $5D$ metric has
signature $(- + + + +)$. 

This is a generalisation of the usual Randall-Sundrum warped metric
ansatz (\cite{RS1, RS2}) to the case of a non-trivial
four-dimensional metric. Such cosmological solutions have been studied
in the context of infinitely thin branes by a number of authors
(\cite{perez-lorenzana, basicRScosmo, padilla, kaloper}). We
shall refer to $e^{- \sigma(y)}$ as the ``warp factor'' associated with
the warped metric, and
$\sigma(y)$ as the ``warp factor exponent''.

This simple ansatz clearly does not encompass all possible 5D metrics
that yield effective 4D gravity. In particular, if the bulk component
$T_{44}$ of the energy-momentum tensor depends only on the bulk
coordinate, then the Einstein equations require that the effective 4D
metric generated by this ansatz must be of constant curvature,
allowing dS$_4$ and AdS$_4$ brane cosmology, but prohibiting more general
FRW solutions.

It is well known that in the case of an infinitely thin brane, the
presence of matter on the brane causes mixing of the bulk and brane
coordinates in the metric, in the coordinate system where
the brane is a hypersurface of constant $y$  (\cite{padilla, brane bending}). This effect is frequently
termed
``brane bending''. A simple concrete example is shown in
\cite{perez-lorenzana}, where an FRW ansatz is used for the 5D
metric, the brane lies at $y = 0$, and the metric components are separable into bulk and brane
coordinates only so long as the brane tension is the sole contributor
to the energy-momentum tensor. However, the ansatz of Eq. \ref{general warped metric} is useful for studying systems where the brane
tension dominates the matter contribution to the energy-momentum
tensor, as in some cases the ensuing Einstein equations can be solved
exactly.

Denoting the Ricci scalar computed from the 5D metric
$g_{M N}$ by $R^{(5)}$, and the 4D Ricci scalar computed from 
$g^{(4)}_{\mu \nu}$ by $R^{(4)}$, we find that the 5D Ricci scalar
and Einstein tensor can be written in terms of their 4D counterparts
and the warp factor as per
\begin{equation} G_{M N} = \pmatrix{ G^{(4)}_{\mu \nu} + 3 g^{(4)}_{\mu \nu} e^{- 2
\sigma(y)} \left( \sigma''(y) - 2 \left( \sigma'(y) \right)^2 \right)
& 0 \cr 0 & - \frac{1}{2} e^{2 \sigma(y)} R^{(4)} - 6 \left(\sigma'(y)
\right)^2}, \label{einstein tensor} \end{equation}
\begin{equation} R^{(5)} = e^{2 \sigma(y)} R^{(4)} + \left( 20 \left( \sigma'(y)
\right)^2 - 8 \sigma''(y) \right). \label{curvature scalar} \end{equation}

The Einstein equations in five dimensions are,
\begin{equation} G_{M N} = - 8 \pi G T_{M N}, \end{equation}
and in particular, writing $8 \pi G = \kappa^2$, we see that
\begin{equation} \kappa^2 T_{44} = \left( \frac{1}{2} e^{2
\sigma(y)} R^{(4)} + 6 \left(\sigma'(y) \right)^2 \right),
\end{equation}
implying that a time-varying 4D curvature scalar is associated with a time-dependent
$T_{44}$  (similarly, variation of the curvature scalar with the other
brane coordinates requires $T_{44}$ to depend on those coordinates). Thus if the brane tension is
independent of the brane coordinates $(t, x^1, x^2, x^3)$, and the trapped matter (if present) does not contribute to
$T_{44}$, only constant-curvature solutions are possible with this
metric ansatz.

Since we expect these four-dimensional metrics to be valid only in the
case where the brane tension dominates over the matter contribution to the energy-momentum
tensor, let us compute the 5D energy-momentum tensor derived from a
scalar field $\Phi$ and associated potential $V(\Phi)$.

In the absence of matter sources, the system is described by the usual
Einstein-Hilbert action with additional scalar field terms
(\cite{weinberg, birrell}),
\begin{eqnarray} S & = & \int d^5 x \sqrt{-g(x)} \left[\frac{-1}{2
\kappa^2} R^{(5)}(x) + \mathcal{L}_{\Phi} \right] \\ \mathcal{L}_{\Phi} & = & - \frac{1}{2} g^{M N} \, \partial_{M} \Phi \,
\partial_{N} \Phi \, - V(\Phi),  \\
g(x) & \equiv & \textrm{Det} \left( g_{M N}
\right). \end{eqnarray}

The resulting energy-momentum tensor is easily computed as,
\begin{eqnarray} T^{M N} & = & 2 \, \frac{\delta \mathcal{L}_{\Phi}}{\delta
g_{M N}} + g^{M N} \mathcal{L}_{\Phi} \nonumber \\ & = & g^{M S} g^{N R} \partial_{R} \Phi \,
\partial_{S} \Phi - g^{M N} \left( V(\Phi) + \frac{1}{2} g^{R S} \, \partial_{R} \Phi \,
\partial_{S} \Phi \right).  \end{eqnarray}

Suppose the scalar field $\Phi$ is a
function of the bulk coordinate $y$ only. Then using the separable
metric ansatz of Eq. \ref{general warped metric}, the non-zero
elements of the energy-momentum tensor become,
\begin{eqnarray} T_{\mu \nu} & = & - e^{- 2 \sigma(y)}
g^{(4)}_{\mu \nu} \left( V(\Phi) + \frac{1}{2} \left( \Phi'(y)
\right)^2 \right), \\ T_{44} & = & \frac{1}{2} \left( \Phi'(y)
\right)^2  - V(\Phi). \end{eqnarray}

The Einstein equations, $G_{M N} = - \kappa^2 T_{M N}$, and Eqs.
\ref{einstein tensor}-\ref{curvature scalar}, then
yield:
\begin{eqnarray} G^{(4)}_{\mu \nu}
& = & e^{- 2 \sigma(y)}
g^{(4)}_{\mu \nu} \left[ \kappa^2 \left( V(\Phi) + \frac{1}{2} \left( \Phi'(y)
\right)^2 \right) - 3  \left( \sigma''(y) - 2 \left( \sigma'(y)
\right)^2 \right) \right], \\  R^{(4)} & = & 2 e^{- 2
\sigma(y) } \left[- 6
\left(\sigma'(y) \right)^2 +  \kappa^2 \left(  \frac{1}{2} \left( \Phi'(y)
\right)^2  - V(\Phi) \right) \right]. \end{eqnarray}

For these equations to be consistent, there must be some constant
$\gamma$ such that $G^{(4)}_{\mu \nu} = \gamma g^{(4)}_{\mu \nu}$
($\Rightarrow R^{(4)} = - 4 \gamma$). Then we obtain the equations of motion,
 \begin{eqnarray} \gamma e^{2 \sigma(y)} & = &\kappa^2 \left( V(\Phi) + \frac{1}{2} \left( \Phi'(y)
\right)^2 \right) - 3  \left( \sigma''(y) - 2 \left( \sigma'(y)
\right)^2 \right), \label{equation of motion 1} \\
& = & 3
\left(\sigma'(y) \right)^2 - \frac{\kappa^2}{2} \left(  \frac{1}{2} \left( \Phi'(y)
\right)^2  - V(\Phi) \right). \label{equation of motion 2} \end{eqnarray}

We can eliminate the potential from these equations and obtain the
result,
\begin{equation}   3 \sigma''(y) - \kappa^2 \left(\Phi'(y)
\right)^2 = \gamma e^{2 \sigma(y)}. \label{V-less equation of motion}
\end{equation}

The potential $V$ can easily be obtained as a function of $y$ from either of Eqs. \ref{equation of
motion 1}-\ref{equation of motion 2},
\begin{equation} V(y) = \frac{1}{2} \left( \Phi'(y) \right)^2 -
\frac{6}{\kappa^2} \left( \sigma'(y) \right)^2 + \frac{2
\gamma}{\kappa^2} e^{2 \sigma(y)}. \label{equation for V} \end{equation} 

In the case of AdS$_4$ space, $\gamma$ is negative, while for dS$_4$
space the converse holds true. Minkowski space corresponds
to the case $\gamma = 0$.

For $\gamma \ne 0$, we can define the dimensionless quantities,
\begin{eqnarray} \eta & = & y \sqrt{\frac{|\gamma|}{3}},
\label{dimensionless y} \\ \hat{\Phi}(\eta)
& = & \frac{\kappa}{\sqrt{3}} \Phi \left( y \sqrt{\frac{|\gamma|}{3}}
\right), \label{dimensionless Phi} \\ \hat{V}(\eta)& = & \frac{\kappa^2}{|\gamma|} V \left( y \sqrt{\frac{|\gamma|}{3}}
\right). \label{dimensionless V} \end{eqnarray}

In terms of these dimensionless quantities, Eqs. \ref{V-less
equation of motion} and \ref{equation for V} become,
\begin{eqnarray} e^{2 \sigma(\eta)} & = & \frac{|\gamma|}{\gamma} \left(
\sigma''(\eta) - \left( \hat{\Phi}'(\eta) \right)^2 \right),
\label{dimensionless equation of motion} \\
\hat{V}(\eta) & = & \frac{1}{2} \left( \hat{\Phi}'(\eta) \right)^2 - 2 \left(
\sigma'(\eta) \right)^2 + 2 \frac{\gamma}{|\gamma|} e^{2
\sigma(\eta)}. \label{dimensionless equation for V} \end{eqnarray}

Note that, independent of the value of $\gamma$, differentiating
Eq. \ref{equation of motion 2} and then substituting Equation
\ref{V-less equation of motion} yields the Klein-Gordon equation for
the scalar field (this relationship between the Einstein equations and the
Klein-Gordon equation is discussed in \cite{csaki+erlich+hollowood+shirman}),
\begin{equation} \frac{\mathrm{d} V}{\mathrm{d} \Phi} = \Phi''(y) - 4
\sigma'(y) \Phi'(y), \label{klein-gordon equation} \end{equation}
or using the dimensionless quantities defined above,
\begin{equation} \frac{\mathrm{d} \hat{V}}{\mathrm{d} \hat{\Phi}} = \hat{\Phi}''(\eta) - 4
\sigma'(\eta) \hat{\Phi}'(\eta). \label{dimensionless klein-gordon
equation} \end{equation}

Later we will wish to discuss configurations where the metric elements
go to zero at some point in the bulk. In this case, $\sigma$ is an
unsuitable parameterisation as it diverges to $\infty$ at zeroes of the warp
factor. If in Eq. \ref{general warped metric} we replace $e^{-2
\sigma}$ with $\omega^2$, where $\omega$ is a real function of the
bulk coordinate, then the equations of motion (Eqs. \ref{V-less
equation of motion}-\ref{equation for V}) can be computed exactly as
previously, and become,
\begin{equation} 3 \left( \omega'(y) \right)^2 - 3 \omega(y)
\omega''(y) - \kappa^2 \left( \Phi'(y) \right)^2 \omega(y)^2 = \gamma, \label{V-less equation of motion for omega}
\end{equation}
\begin {equation} V(y) = \frac{1}{2} \left( \Phi'(y) \right)^2 -
\frac{6}{\kappa^2} \left( \frac{\omega'(y)}{\omega(y)} \right)^2 + \frac{2
\gamma}{\kappa^2 \omega(y)^2}. \label{equation for V in omega}
\end{equation}
(Note: It immediately follows from these equations of motion that if
the first and second derivatives of $\omega$ with respect to the bulk coordinate
remain finite, then $\omega$ may possess zeroes only if $\gamma \ge
0$, i.e. where the 4D metric is dS$_4$ or Minkowski. We will later
show that zeroes in $\omega$ inevitably arise for the case of a smooth
$\omega$ with dS$_4$ brane cosmology.)

In terms of the dimensionless quantities defined in
Eqs. \ref{dimensionless y}-\ref{dimensionless V}, the equations of
motion for $\omega$ are,
\begin{eqnarray} \frac{|\gamma|}{\gamma} & = &
\left( \omega'(\eta) \right)^2 - \omega''(\eta) \omega(\eta) -
\omega(\eta)^2 \left( \hat{\Phi}'(\eta) \right)^2,
\label{dimensionless equation of motion in omega} \\
\hat{V}(\eta) & = & \frac{1}{2} \left( \hat{\Phi}'(\eta) \right)^2 - 2 \left(
\frac{\omega'(\eta)}{\omega(\eta)} \right)^2 + 2 \frac{\gamma}{|\gamma|} 
\frac{1}{\omega(\eta)^2}. \label{dimensionless equation for V in omega}
\end{eqnarray}

We will preferentially use the $\sigma$ notation when not dealing directly
with cases where $\sigma \rightarrow \infty$, for ease of
comparison with the literature. Coordinate singularities in the metric, where $\sigma$
diverges (and $\omega$ goes to zero), correspond to bulk horizons, and
most previous braneworld studies have been concerned only with the behaviour
of the brane-bulk system in the region bounded by these horizons.  

\section{Scalar field kink configurations supporting a Minkowski brane}

\label{scalar field kink solutions for a Minkowski brane}

In the case where $\gamma = 0$ and the brane metric corresponds to
Minkowski space, Eq. \ref{V-less equation of motion} can be
integrated to give a warp factor corresponding to any chosen scalar
field profile. The potential generating this scalar field can then be
analytically determined from Eq. \ref{equation for V} with
$\gamma = 0$. Many solutions to the coupled gravity-scalar system for
a Minkowski brane have been discussed in the literature (see for
example \cite{scalar field solutions, kehagias+tamvakis, bronnikov+meierovich,
davidson+mannheim, koley+kar1, koley+kar2}).

Two simple examples are the kink configurations $\Phi_1(y) = ( \sqrt{3}
A / \kappa ) \tanh (r y)$,
$\Phi_2(y) = ( \sqrt{3} A / \kappa ) \arctan \sinh (r y)$, where $A$
is a dimensionless constant and $r$ has dimensions of inverse length. Eq. \ref{V-less equation of
motion} then yields the corresponding warp factor exponents $\sigma_1, \sigma_2$:
\begin{eqnarray} \sigma_1(y) & = & A^2 \left( \frac{2}{3}\ln \cosh (r y) - \frac{1}{6}
\textrm{sech}^2 (r y) + B r y + C \right), \\  \sigma_2(y) & = & A^2
\left( \ln \cosh (r y) + B r y + C \right), \end{eqnarray}
where $B$ and $C$ are dimensionless constants of integration. If
the warp factor exponent is an even function of $y$ (a usual
symmetry of domain-wall solutions and the RS model with a single brane
\cite{RS2}), then it follows that the integration constant $B$ must be
zero. Then in both cases, the warp factors $e^{- \sigma}$ converge to zero far from the
brane, like the warp factor of \cite{RS2} for the case of a fine-tuned
infinitely thin brane. The corresponding potentials $V_1, V_2$ can easily be obtained from
Eq. \ref{equation for V}: $V_1$ is a sextic polynomial in $\Phi$
(\cite{kehagias+tamvakis}), while $V_2(\Phi)$ is
\begin{eqnarray} V_2(\Phi) & = & - \frac{3 A^2}{2} \frac{r^2}{\kappa^2} \left[ \left( 4 A^2
+ 1 \right) \sin^2\left(\frac{\kappa}{A \sqrt{3}} \Phi \right) + 8 A^2 B \sin \left( \frac{\kappa}{A \sqrt{3}} \Phi \right) + \right. \nonumber \\
& & \left. + \left( 4 A^2
B^2 - 1 \right)  \right]. \end{eqnarray}
If $B$ vanishes (required if the warp factor is even) then the
potential is also an even function of $\Phi$. 

This method is useful for deriving analytically tractable solutions, but is conceptually somewhat unsatisfactory because the
system is fundamentally determined by the potential, not the profile
of the scalar field. ``Superpotential'' methods have been used to
reduce the equations of motion to a system of first-order equations by
introducing an auxiliary potential
(\cite{dewolfe+freedman+gubser+karch, afonso, flanagan+tye+wasserman2}), allowing the scalar field profile and metric
corresponding to a wide range of potentials to be derived. These
calculations can often be performed analytically, although writing the
potential as a function of the scalar field rather than the bulk
coordinate requires an inversion of the scalar field profile which is
often analytically intractable.

If $\gamma$ is non-zero, i.e. the 4D metric has non-zero constant
curvature, then the problem becomes significantly more complex and this
superpotential approach no longer yields easily solvable
equations of motion. This problem has been investigated in
\cite{dewolfe+freedman+gubser+karch, afonso, flanagan+tye+wasserman2}).

A modified superpotential method can be employed to generate
scalar field configurations and metric warp factors corresponding to a given
auxiliary potential (\cite{afonso}). However, it is applicable only to a limited
class of potentials. The solutions found by this method with AdS$_4$
brane cosmology exhibit kink-like scalar field configurations, but the
dS$_4$ solutions contain naked curvature singularities in the metric,
associated with infinities
in the scalar field as the potential becomes unbounded below, at a finite
distance from the brane.

\section{Properties of solutions to the Einstein equations with dS$_4$ cosmology}

\label{properties of solutions to the Einstein equations with dS$_4$ cosmology}

Putting the superpotential approach aside, let us consider the general
properties of solutions to the equations of motion
(Eqs. \ref{dimensionless equation of motion}-\ref{dimensionless
equation for V}) when $\gamma > 0$. We
will prove that in this case, corresponding to dS$_4$ brane cosmology,
there are no
smooth, even solutions for the exponent $\sigma(\eta)$: if the warp
factor $e^{- \sigma(\eta)}$ is to be smooth and even, it must
possess zeroes.

We begin by noting that for $\gamma > 0$, $\sigma''(\eta) = (
\hat{\Phi}'(\eta) )^2 + e^{2 \sigma(\eta)} > 0 \,
\forall \, \eta$. Thus $\sigma'(\eta)$ is monotonically
increasing $\forall \, \eta$. Suppose that $\sigma(\eta)$ is an even continuous function of $\eta$
with continuous and well-defined first and second derivatives: then $\sigma'(0) = 0$ and
consequently $\sigma'(\eta)$ is negative for $\eta < 0$ and positive for $\eta
> 0$. Thus if $\sigma(\eta)$ is a smooth function of $\eta$, then it must be
strictly decreasing for $\eta < 0$ and strictly increasing for $\eta > 0$,
and it follows that $\sigma(0)$ is a minimum of the metric exponent $\sigma$: $\sigma(\eta) > \sigma(0)
\, \forall \, \eta \ne 0$.

Substituting back into Eq. \ref{dimensionless equation of motion}, it
follows that 
\begin{equation} \sigma''(\eta) \ge  e^{2 \sigma(\eta)} \ge  e^{2 \sigma(0)}, \end{equation}
and integrating this inequality implies that $\sigma(\eta), |\sigma'(\eta)| \rightarrow
\infty$ as $\eta \rightarrow \pm \infty$.

Let $g(\eta) \equiv \sigma'(\eta) e^{- \sigma(\eta)}$. Then we can rewrite
Eq. \ref{dimensionless equation of motion} in the form,
\begin{equation} g'(\eta) e^{- \sigma(\eta)} + g(\eta)^2 = 1 +
 e^{- 2 \sigma(\eta)} \left( \hat{\Phi}'(\eta) \right)^2. \end{equation}
By hypothesis, $g(\eta)$ is a smooth odd function of $\eta$, with $g(\eta) > 0$
for $\eta > 0$, and $g'(0) = \sigma''(0) e^{- \sigma(0)} > 0$.

Now wherever $g'(\eta) \le 0$, we must have $g(\eta)^2 \ge 1$, which
for $\eta > 0$ implies $g(\eta) \ge 1$. Taking the
contrapositive, in the range $\eta > 0$, whenever $g(\eta) <
1$, $\Rightarrow g'(\eta) > 0$. Thus if $g(\eta) < 1, \, \, \forall \, \eta > 0$, then $g$ is bounded and
monotonically increasing and converges to some (strictly) positive
real number $\delta \le 1$. Furthermore, since $g(\eta)$ is
monotonically increasing on any interval for which $g(\eta) <
1$, and $g$ is continuous, it follows that if $\exists
\, \eta_0$ s.t. $g(\eta_0) \ge 1$, then $\eta \ge \eta_0 \Rightarrow
g(\eta) \ge 1$.

In both of these cases, $\exists$ some real and strictly positive
number $\epsilon$, and some $\eta_0 > 0$, such that $\eta \ge \eta_0 \Rightarrow
g(\eta) \ge \epsilon$. (In the first case, we can simply choose $\epsilon =
\delta / 2$.)

But then choosing some $\eta_1 > \eta_0$, and integrating $g(\eta)$ over this
interval, we find,
\begin{equation}  \epsilon (\eta_1 - \eta_0) \le \int_{\eta_0}^{\eta_1} g(\eta) \, \mathrm{d} \eta  =  \int_{\eta_0}^{\eta_1}
\sigma'(\eta) e^{- \sigma(\eta)} \, \mathrm{d} \eta  = - e^{- \sigma(\eta_1)} + e^{-
\sigma(\eta_0)} \le e^{- \sigma(\eta_0)} \le e^{- \sigma(0)}. \end{equation}
But this is a contradiction, as by choosing $\eta_1$ arbitrarily large we
can obtain an arbitrarily large lower bound on $e^{- \sigma(0)}$,
which must be finite.

Consequently, there is no even, smooth warp factor exponent $\sigma(y)$ defined on the
whole real line that yields a four-dimensional deSitter
cosmology. It is possible that the warp factor $\omega(y)$ is smooth,
but in this case it must contain zeroes (as smoothness of $\omega =
\pm e^{- \sigma}$
implies smoothness of $\sigma$ except where $\omega$ changes sign).

It is well known that warped metrics with dS$_4$ slices often contain
horizons or naked singularities at a finite distance from the brane
(\cite{kaloper, dewolfe+freedman+gubser+karch, gremm, davidson+mannheim, karch+randall}). The properties of horizons
in the case of a thin brane with no scalar field have been discussed
in  \cite{kaloper, flanagan-et-al}, and curvature
singularities in the presence of a scalar field were considered in
\cite{flanagan+tye+wasserman2, gremm}. We have demonstrated the strong result that
such features are inevitable in the case of a
brane supported by a scalar field.

In the simplest analytic solutions for this system, the scalar field
diverges to infinity at these points, and naked curvature
singularities occur (\cite{afonso, flanagan+tye+wasserman2, gremm}). We will show that it is possible to avoid a diverging curvature at
these points, and to obtain a kink-like configuration for the scalar
field supporting the dS$_4$ brane, but at present we have only obtained a partially
analytic example of such a solution.

The physical interpretation of singularities in the metric,
for a model where gravity is coupled to a scalar field,
has been discussed in the literature (\cite{gremm, davidson+mannheim}). In particular, Gremm
suggests that five-dimensional spacetimes consisting of domain walls interpolating between spaces with naked
curvature singularities may be interpreted as four-dimensional gravity coupled
to a non-conformal field theory  \cite{gremm}. Gremm also claims to present a
warped metric with dS$_4$ slices where the curvature remains finite at
the metric zeroes, but the scalar field and potential
diverge, but we have not been able to verify this solution. Davidson and Mannheim give a similar solution in the case of
an infinitesimally thin brane, with a divergent scalar field and potential but bounded
curvature at the metric zeroes \cite{davidson+mannheim}, and suggest
that such a solution might provide a mechanism for dynamical
compactification of the extra dimension.

To distinguish between curvature and coordinate singularities, we
examine the five-dimensional curvature scalar, given by
Eq. \ref{curvature scalar}. Also, $- \kappa^2 T^{M}_{M} = G^{M}_{M} = R^{M}_{M} -
(1/2) g^{M}_{M} R = R (1 - n/2)$, where $n$ is the number of
dimensions. So the five-dimensional curvature scalar can be expressed
in terms of the trace of the energy-momentum tensor,
\begin{equation} R^{(5)} = \frac{2 \kappa^2}{3} T^{M}_{M}. \end{equation} 
In the case where the energy-momentum tensor is generated entirely by
the scalar field and potential,
\begin{eqnarray} T^{M}_{M} & = & -4 \left( V(\Phi) + \frac{1}{2}
\left( \Phi'(y) \right)^2 \right) + \frac{1}{2} \left( \Phi'(y)
\right)^2 - V(\Phi) \nonumber \\ & = & - 5 V(\Phi) - \frac{3}{2} \left( \Phi'(y)
\right)^2. \label{energy-momentum trace} \end{eqnarray}
Thus, the curvature scalar can only become singular if the potential, the
scalar field derivative, or both, diverge to infinity. If the
potential and scalar field remain bounded at a zero in the metric,
this is sufficient (albeit not necessary, as diverging terms on the RHS of
Eq. \ref{energy-momentum trace} may cancel each other out) to ensure that the zero is
not associated with a curvature singularity.

In the case where $\gamma \ne 0$, we can also write,
\begin{eqnarray} R^{(5)} & = &  e^{2 \sigma(\eta)} R^{(4)} +
\frac{|\gamma|}{3} \left( 20 \left( \sigma'(\eta)
\right)^2 - 8 \sigma''(\eta) \right) \\ & = & - |\gamma| \left( \frac{10}{3} \hat{V}(\hat{\Phi}(\eta)) + \left( \hat{\Phi}'(\eta)
\right)^2 \right). \label{5D curvature scalar - dimensionless}
\end{eqnarray}

\section{Partially analytic solution with dS$_4$ slices and no
curvature singularities}

\label{Partially analytic solution with dS$_4$ slices and no curvature singularities}

Having demonstrated above that the dS$_4$ case must produce metric zeroes
(or a divergent metric) in the bulk, to proceed with our analysis of that case we now no
longer take the warp factor to be the exponential of a 
real-valued function.  Instead, we work directly with the warp factor
$\omega$ introduced previously, employing
Eqs. \ref{dimensionless equation of motion in omega}-\ref{dimensionless equation for V in omega}
as the equations of motion.

Let us denote the roots of $\omega(\eta)$ by $\eta_0$,
i.e. $\omega(\eta_0)=0$. Then in the dS$_4$ case where $\gamma$ is
positive, it follows immediately from Eq. \ref{dimensionless equation of motion in omega} that
\begin{equation} \omega'(\eta_0) =
\pm 1, \label{condition1} \end{equation} 
for the case of interest where $\omega''$ and $\hat{\Phi}'$ remain finite
at the metric zeroes. Differentiating both sides
of the equation of motion up to fourth order and evaluating at $y=y_0$, we obtain the
relations:
\begin{eqnarray} \omega''(\eta_0) & = & 0, \label{condition2} \\
\hat{\Phi}'(\eta_0) & = & 0, \label{phi-flat-condition} \\
\omega^{(4)}(\eta_0) & = & 0. \label{condition3} \end{eqnarray}
(See also \cite{flanagan+tye+wasserman2} for an alternate approach
yielding Eq. \ref{phi-flat-condition}).

If $\hat{\Phi}(\eta)$ is not $1:1$, then it cannot be inverted to yield an expression for
$\hat{V}$ as a function of $\hat{\Phi}$, and thus $\hat{V}$ may fail to be a single-valued
function of $\hat{\Phi}$. It is still \emph{possible} that $\hat{V}$ might be a
single-valued function of $\hat{\Phi}$ in any case, but we can ensure
this is true by requiring $\hat{\Phi}$ to
be nondecreasing. With this additional constraint, any point where
$\hat{\Phi}'(\eta)=0$ must also be a turning point in the derivative,
i.e. $\hat{\Phi}''(\eta)$ must also be zero.

This assumption has consequences for the form of the required
potential. We can rewrite the Klein-Gordon equation (Eq.
\ref{dimensionless klein-gordon equation}) in terms of $\omega(\eta)$, 
\begin{equation} \frac{\mathrm{d} \hat{V}}{\mathrm{d} \hat{\Phi}} = \hat{\Phi}''(\eta) + 4
\frac{\omega'(\eta)}{\omega(\eta)}
\hat{\Phi}'(\eta). \label{dimensionless klein-gordon equation 2} \end{equation}
If $\hat{\Phi}(\eta)$ is nondecreasing and continuous, then
$\hat{\Phi}'(\eta) \ge 0 \, \forall \, \eta \,$, and as zeroes in the metric
correspond to points where $\omega'(\eta) / \omega(\eta)$ changes sign, it
follows that zeroes in the metric are either zeroes in the derivative of
the potential with respect to $\hat{\Phi}$ (if $\mathrm{d} \hat{V} / \mathrm{d} \hat{\Phi}$ is
continuous at these points), or cusps where $\mathrm{d} \hat{V} /
\mathrm{d} \hat{\Phi}$ changes sign. This latter case was proposed as a
possibility by Davidson and Mannheim \cite{davidson+mannheim}.

Imposing the condition that
$\hat{\Phi}''(\eta_0) = 0$ and differentiating the equation of motion up to
sixth order yields the further relations:
\begin{eqnarray} \left( \omega^{(3)}(\eta_0) \right)^2 & = &
\omega^{(1)}(\eta_0) \omega^{(5)}(\eta_0), \label{condition4} \\
\omega^{(6)}(\eta_0) & = & 0 \, .
\label{condition5} \end{eqnarray}

Note that the conditions on the metric are a result of our somewhat artificial approach in
starting from the metric and attempting to reconstruct a physically
reasonable scalar field kink and potential. It is not clear to what
degree they indicate constraints on the class of potentials that can give rise
to a scalar kink supporting a dS$_4$ domain wall. However, we have
also derived constraints that apply directly to the potential,
implying that it must have turning points or points of inflection at
the zeroes of the metric. This constitutes a limitation on the physical
potentials that can give rise to a smooth singularity-free scalar
field and metric with dS$_4$ cosmology on the brane. In flat space we would
expect $\hat{V}(\hat{\Phi})$ to
possess minima at the asymptotic values of the scalar field $\hat{\Phi}(\eta)$,
that is at the limits $\lim_{\eta \rightarrow \pm \infty} \hat{\Phi}(\eta)$, but,
as noted above, in the
present case the potential must
have extrema at the zeroes of the metric.
The values of
$\hat{\Phi}(\eta)$ at the zeroes of the
metric will generally not coincide with the asymptotes, so a simple
double-well potential cannot give rise to the desired metric and
scalar field configuration.

A straightforward but inelegant method for writing down a warp factor satisfying
these conditions is to start with a simple $\cosh$ function,
motivated by the warp factor in the case of an infinitely thin
brane with a time-dependent metric (\cite{perez-lorenzana,
basicRScosmo, karch+randall}). Adding additional linearly independent even
functions (sech, sech$^2$ and sech$^4$ functions are suitable) 
which decrease rapidly far from the brane and are zero at
the roots of the warp factor, we retain the
essential features of the delta-function brane solution, but the
coefficients of these terms can be adjusted to fine-tune the
derivatives of the metric as required.

Once the metric has been written down, Eq. \ref{dimensionless
equation of motion in omega} immediately yields the derivative of the dimensionless scalar
field $\hat{\Phi}'(\eta)$, and integrating gives $\hat{\Phi}(\eta)$ itself. Equation
\ref{dimensionless equation for V in omega} then yields the dimensionless potential $\hat{V}$ as a
function of $\eta$, and inverting the function $\hat{\Phi}(\eta)$ yields an
expression for $\hat{V}$ as a function of $\hat{\Phi}$.

However, at this stage the integration to obtain $\hat{\Phi}(\eta)$ (and
consequently the inversion of this function to obtain $\hat{V}(\hat{\Phi})$) has
not been performed analytically, due to the inelegance of our method
for writing down a metric satisfying all the requirements of the
problem.

Specifically, the metric ansatz,
\begin{eqnarray} \omega(\eta) & = & A - R \cosh(r \eta) + B \left( \textrm{sech}(r \eta)
- \frac{R}{A} \right) + C \left( \textrm{sech}^2 ( r \eta) - \left(
  \frac{R}{A} \right)^2 \right) + \nonumber \\ & & +  D \left( \textrm{sech}^4 (r \eta) -
  \left( \frac{R}{A} \right)^4 \right) + E \left( \textrm{sech}^6 (r
  \eta) - \left( \frac{R}{A} \right)^6 \right),
  \label{bruteforce-metric-ansatz} \end{eqnarray}
satisfies Eqs. \ref{condition1}-\ref{condition2} and
  \ref{condition3}-\ref{condition4}, with $A$ and $R$ as
  adjustable parameters and $B$, $C$ and $D$ given by,
\begin{eqnarray} B & = &  \left\{ A^{9} \left( 16 A^{10} - 200 R^2 A^{8} + 1294 R^4 A^{6} -
  3327 R^6 A^{4} + 3612 R^8 A^{2} - 1400 R^{10} \right) +
  \right. \nonumber \\ & & \left. + E R^6
  \left( - 8192 A^{12}  + 73728 A^{10}
  R^2  - 279040 A^8 R^{4} +567296 A^6 R^{6} -  \right. \right. \nonumber \\ & & \left. \left.
  -649152 A^4 R^{8} + + 393792 A^2 R^{10} - 98448 R^{12} \right) \right\}
  / \left[ A^5 R \left(  16 A^{12}  -168  A^{10} R^2 + \right. \right. \nonumber \\ & & \left. \left. +  766 A^8 R^4 -  
  2019  A^6 R^6 + 3070  A^4
  R^8 -2432  A^2
  R^{10} + 768 R^{12} \right) \right], \label{Bsolution}
  \\ C & = & \frac{1}{2} \left\{ A^9 \left( 28 A^{8} - 443 R^2 A^{6} + 1476 R^4
  A^{4}  -1840 R^6 A^{2} + 
  784 R^8  \right) +  E R^4 \left( 5600 A^{12}-
  \right. \right. \nonumber \\  & & \left. \left. -53648 A^{10} R^2 + 217828 A^8 R^4 -479506 A^6 R^6 +595020 A^4
  R^8 -389424 A^2 R^{10} + \right. \right. \nonumber \\ & & \left. \left. +104160 R^{12} \right) \right\} / \left[ A^4
  \left( 16 A^{12} - 168  A^{10} R^2 + 766 A^8 R^4  - 2019 A^6 R^6 +
  3070 A^4 R^8 - \right. \right. \nonumber \\ & & \left. \left. - 2432 A^2 R^{10} + 768
  R^{12} \right) \right], \label{Csolution} \\ D & = &
  \frac{1}{4} \left\{ A^9 \left( 2 A^{8} + 13 R^2 A^{6}  -80 R^4 A^{4} +
  128 R^6 A^{2} - 64 R^8 \right) + E  R^4 \left(  -896 A^{12} +
  \right. \right. \nonumber \\ & & \left. \left. + 9056
  A^{10} R^2 - 39864 A^8 R^4  + 98164 A^6 R^{6}  - 137328 A^4
  R^{8} + 100416  A^2 R^{10} - \right. \right. \nonumber \\ & &
  \left. \left.  -29568 R^{12} \right) \right\} / \left[ R^2 A^2 \left( 16 A^{12} - 168  A^{10} R^2 + 766 A^8 R^4  - 2019 A^6 R^6 +
  3070 A^4 R^8 -  \nonumber \right. \right. \\ & &
  \left. \left. - 2432 A^2 R^{10} + 768
  R^{12} \right) \right]. \label{Dsolution} \end{eqnarray}
The parameter $E$ is obtained by solving the quadratic equation,
\begin{eqnarray} 0 & = & E^2 \times (9408 R^{12} A^2 - 11424 R^{14}) +
  \nonumber \\ & & +
E \times (-1225 R^4 A^{11} + 5116 B R^7 A^7 - 2048 C R^6 A^8 - 8112 C R^{10}
A^4 +5488 A^9 R^6  - \nonumber \\ & & - 5040 A^7 R^8 - 1225 B R^5 A^9 +12832 D R^{10} A^4 -
1600 D R^8 A^6 - 4488 B R^9 A^5 - \nonumber \\ & & - 13440 D R^{12} A^2 + 9056 C R^8 A^6)
+ \nonumber \\ & &
+ 1600 D^2 R^8 A^6 +8 B^2 R^2 A^{12}- 3 B R A^{13} C + 72 A^{13} C R^2
  + 1000 A^{11} D R^4 - \nonumber \\ & & - 20 A^{12} B R^3 - 1120 A^9 D R^6 - 3 C A^{15}
  - 150 D R^2 A^{13} - 120 A^{11} C R^4 + \nonumber \\ & & +76 B R^3 A^{11} C - 2080 D^2
  R^{10} A^4 -  144 C^2 R^6 A^8 + 8 A^{14} B R - 960 B R^7 A^7 D -
  \nonumber \\ & & - 112
  B R^5 A^9 C+912 B R^5 A^9 D-150 B R^3 A^{11} D  +
  1568 C R^6 A^8 D  + 96
  C^2 R^4 A^{10} - \nonumber \\ & &  -1760 C R^8 A^6 D-192 C R^4 A^{10} D -14 B^2 R^4
  A^{10}. \label{Esolution} \end{eqnarray}
The parameter $r$ is then determined by solving the equation,
\begin{equation} 1 = \left(\omega'(\eta_0) \right)^2 = \frac{r^2 (A^2 - R^2) (A^7 + B R
  A^5 + 2 C R^2 A^4 + 4 D R^4 A^2 + 6 E
  R^6)^2}{A^{14}}. \label{rsolution} \end{equation}

For example, setting $A=5$, $R=1$, we obtain,
\begin{equation} B \approx 24.0, C \approx 2.53, D \approx 8.17, E
\approx -1.66, r \approx 0.101. \label{sampleparameters} \end{equation}
Figures \ref{dS-bruteforce-metric} and \ref{dS-bruteforce-kink}
illustrate the bulk profile of the metric and dimensionless scalar
field in this case, while Fig. 
\ref{dS-bruteforce-potential} depicts the
required potential.

\FIGURE{
\includegraphics[angle=270,width=0.6\textwidth]{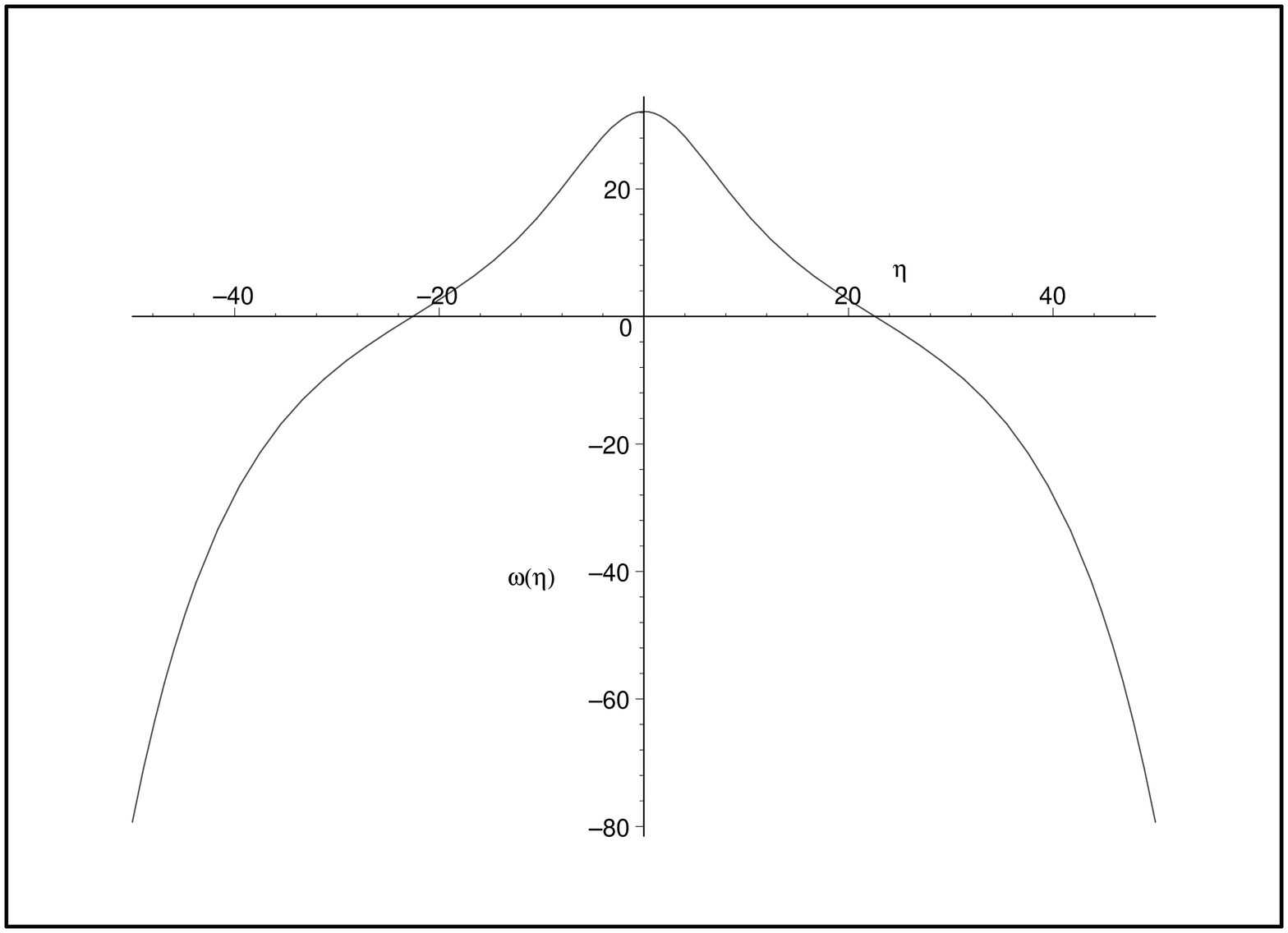}
\caption{Warp factor $\omega(\eta)$ for the smooth kink with dS$_4$
brane cosmology; $A=5,R=1$, see Eqs. \ref{bruteforce-metric-ansatz}-\ref{sampleparameters}.
}
\label{dS-bruteforce-metric}}

\FIGURE{
\includegraphics[angle=270,width=0.6\textwidth]{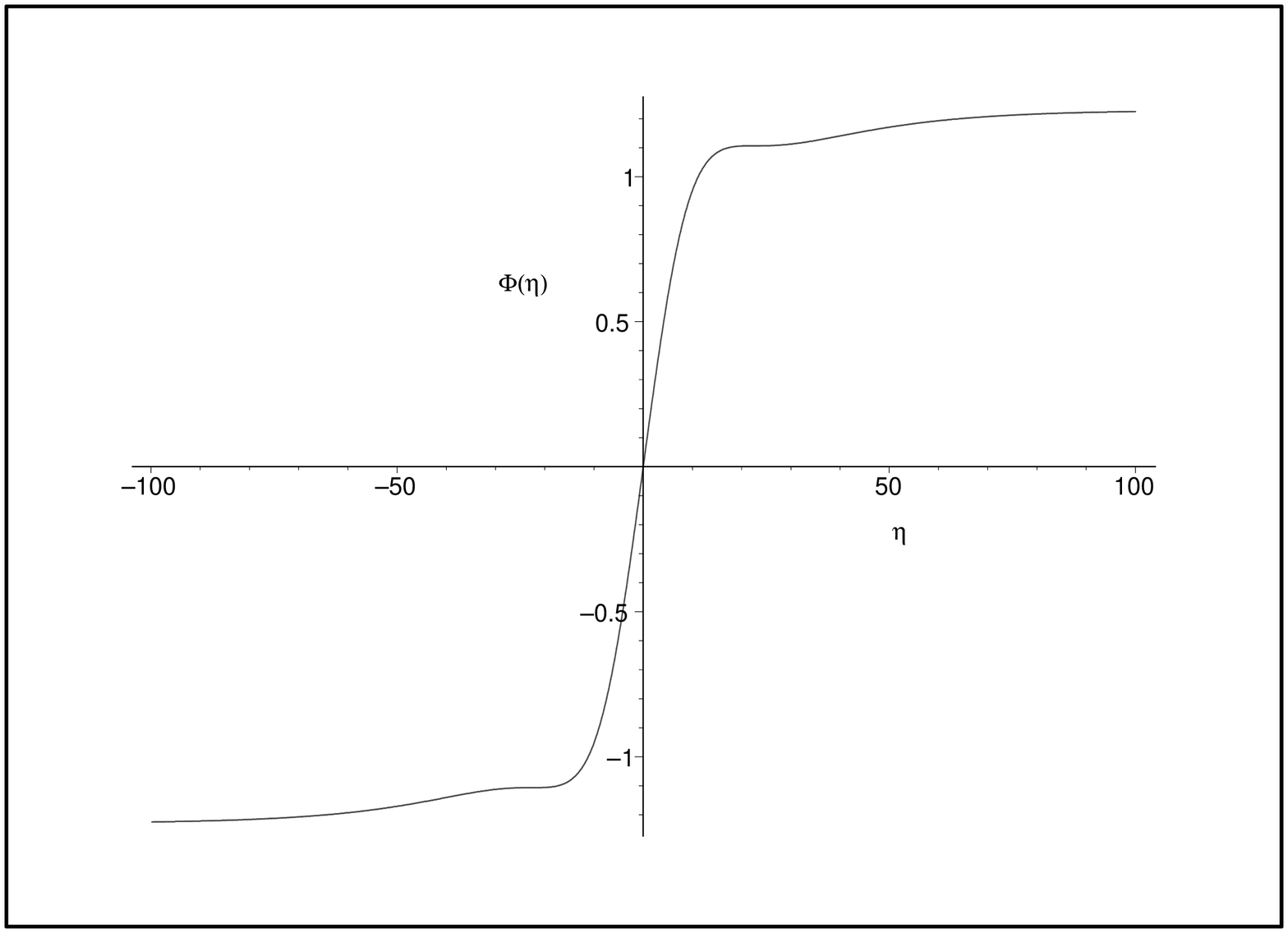}
\caption{Dimensionless scalar field profile $ \hat{\Phi}(\eta)$ generating dS$_4$ brane cosmology;
$A=5,R=1$, see Eqs. \ref{bruteforce-metric-ansatz}-\ref{sampleparameters}.}
\label{dS-bruteforce-kink}}

\FIGURE{
\includegraphics[angle=0,width=0.6\textwidth]{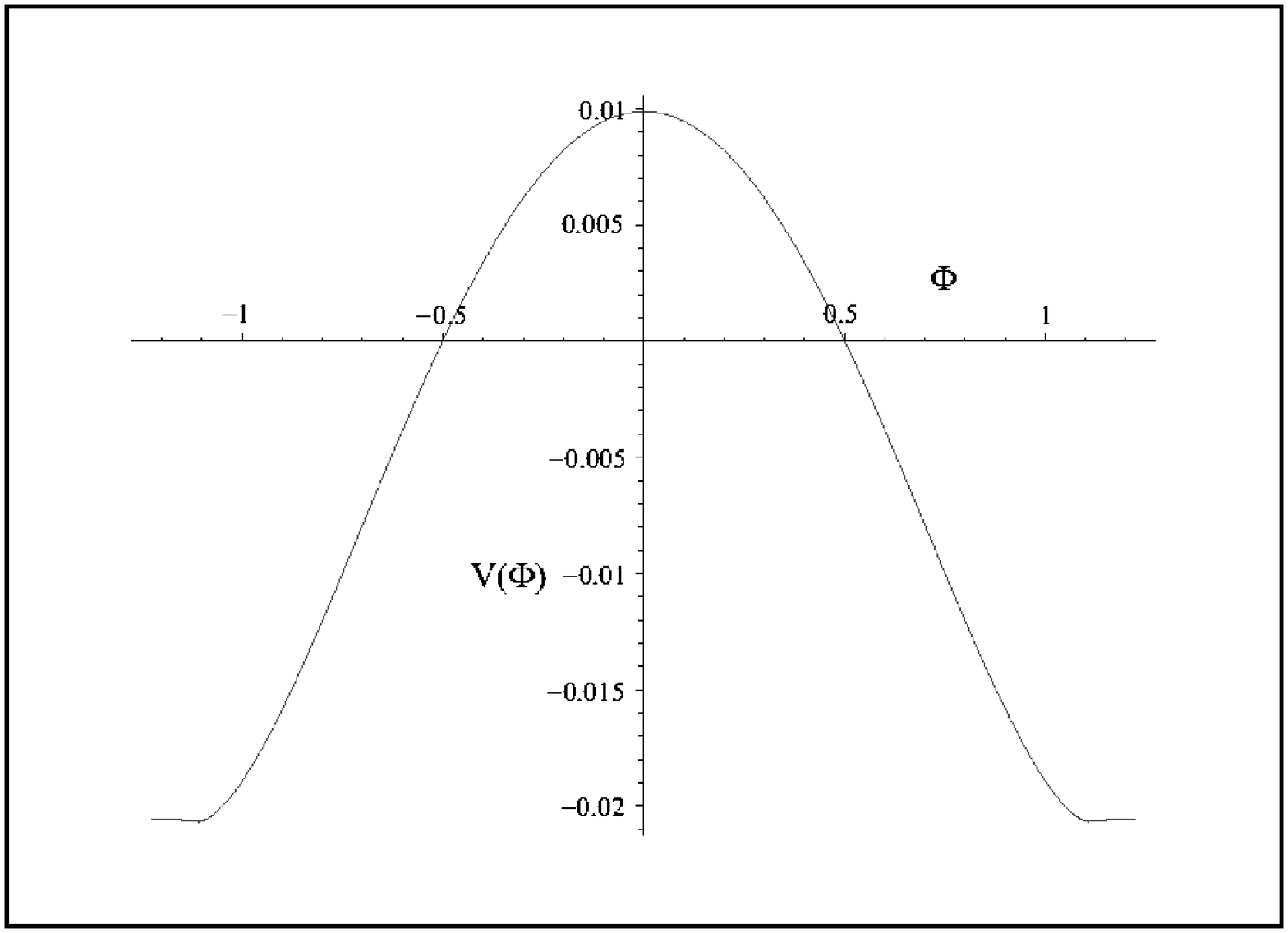}
\caption{Potential $V(\Phi)$ generating dS$_4$ brane cosmology;
$A=5,R=1$, see Eqs. \ref{bruteforce-metric-ansatz}-\ref{sampleparameters}.}
\label{dS-bruteforce-potential}}

\section{``Locally localised gravity'' solutions with AdS$_4$ and dS$_4$ brane cosmology}

\label{locally localised gravity}

In the case where $\gamma < 0$ and the brane cosmology is
anti-deSitter, the metric zeroes which complicate the dS$_4$ case do
not occur and it is possible to write down a fully analytic
solution system. The simplest such solution (which may also be obtained by a superpotential
approach \cite{afonso} starting with a trigonometric periodic
potential) has the warp factor exponent $\sigma(\eta) = A - \ln \cosh
(r \eta)$. However, we shall first consider a slightly more complicated trial warp
factor, and recover this simple solution as a special case. 

Karch and Randall (\cite{karch+randall}) discussed warp factors giving
rise to localised four-dimensional gravity in the case of an
infinitely thin brane. In the case of a brane with AdS$_4$ cosmology,
the warp factor grew exponentially far from the brane, like the
simple solution described above, or like the warped metric presented in
\cite{koley+kar2} generated by a ``ghost'' scalar field. However, close to the brane the metric behaved
qualitatively like
the decreasing warped metric associated with a Minkowski brane
(Sec. \ref{scalar field kink solutions for a Minkowski brane},
\cite{RS2}), and it was this local behaviour that
was responsible for confining gravity to the brane. It seems
reasonable, then, that a similar but smooth metric (with a decreasing
warp factor
close to the brane that then turns around and increases exponentially)
might be required to localise four-dimensional gravity on a scalar
field domain wall in the AdS$_4$ case. An example of such a warp
factor exponent is
plotted in Fig. \ref{localAdSmetric}, and graviton confinement in coupled metric-scalar
systems of this form was studied by Kobayashi et al \cite{soda}.

It is possible to ``smooth out'' warp
factors initially derived for a thin-brane system, essentially by replacing $|\eta|$ with $\ln(2 \cosh \eta)$. Using
this idea as motivation, we can construct exact analytic solutions for
the scalar field and metric with
AdS$_4$ and dS$_4$ slices: the supporting scalar field is kink-like in
the AdS$_4$ case but contains curvature singularities in the dS$_4$
case, as discussed above.

We consider a trial warp factor of the form,
\begin{equation} \omega(\eta) = A \cosh (r \eta) + B \textrm{sech} (r
\eta). \label{locally localised gravity ansatz} \end{equation}
If $0 < A / B < 1$, then this metric has  off-brane turning points
similar to those in the solutions of \cite{karch+randall, soda} (see Fig. \ref{localAdSmetric}). The limit
where $B=0$ corresponds to the simple solution mentioned at the start
of this section. In any case, Eq. \ref{dimensionless equation of
motion in omega} becomes:
\begin{eqnarray} \left(\hat{\Phi}'(\eta)
\right)^2 & = & \frac{1}{\omega(\eta)^2} \left( \left(\omega'(\eta)\right)^2 - \omega''(\eta)
\omega(\eta) - \frac{\gamma}{|\gamma|} \right) \nonumber
\\ & = & \frac{1}{\left(A \cosh(r \eta) + B \textrm{sech}(r \eta)\right)^2}
\times \\ & & \times
\left[ -\frac{\gamma}{|\gamma|} + r^2 \left( -A(A + 4 B) + 4 A B
\textrm{sech}^2(r \eta) + B^2 \textrm{sech}^4 (r \eta) \right)
\right]. \label{kink equation localised gravity} \end{eqnarray}

To ensure that the RHS of the equation is always positive
and to facilitate an analytic solution, we  impose the relation
\begin{equation} - \frac{\gamma}{|\gamma|} = r^2 \left( 4 A^2 + A( A + 4 B)
\right). \label{condition on parameters} \end{equation}
 Eq. \ref{kink equation localised
gravity} then yields,
\begin{equation} \left(\hat{\Phi}'(\eta)
\right)^2  =  r^2 \textrm{sech}^2 (r \eta) \left( \frac{2 A + B \textrm{sech}^2 (r \eta)}{A + B \textrm{sech}^2(r \eta)} \right)^2, \end{equation}
and taking the square root of both sides and integrating yields the
scalar field profile,
\begin{equation} \hat{\Phi}(\eta) = \pm \left[ \arctan
\sinh (r \eta) + \sqrt{\frac{A}{A+B}} \arctan \left( \sqrt{\frac{A}{A+B}}
\sinh (r \eta) \right) \right]. \label{scalar field localised gravity} \end{equation}

The dimensionless potential $\hat{V}$ can easily be written as a function of $\eta$,
\begin{eqnarray} \hat{V}(\eta) & = & \left( \frac{r}{\left( A +
B \textrm{sech}^2 ( r \eta ) \right)} \right)^2 \times \bigg[ - 2 A^2 - 2
A (3 A + 2 B) \textrm{sech}^2 ( r \eta) - \nonumber \\ & &  \hphantom - 2 B (A + B)
\textrm{sech}^4 (r \eta) + \frac{5}{2} B^2 \textrm{sech}^6 (r \eta)
\bigg], \label{locally localised V} \end{eqnarray} 
however $\hat{\Phi}(\eta)$ cannot easily be inverted analytically, so $\hat{V}$ can generally not
be written analytically in terms of standard functions of $\hat{\Phi}$ (although $\hat{\Phi}$ is
$1:1$ and therefore always invertible, so $\hat{V}$ can always be expressed numerically
as a well-defined function of $\hat{\Phi}$). 

AdS$_4$ (dS$_4$) brane cosmology corresponds to the case where $\gamma
< 0$ ($\gamma > 0$),
and by Eq. \ref{condition on parameters}, this is equivalent to
requiring,
\begin{eqnarray} 5 A^2 + 4 A B & > & 0, \, \, \textrm{AdS case},
\label{AdS consistency condition} \\ 5
A^2 + 4 A B &
< & 0, \, \,
\textrm{dS case}.  \label{dS consistency condition} \end{eqnarray}

In the AdS case, Eq. \ref{AdS consistency condition} can obviously be easily satisfied
provided $A, B > 0$. In this case $\sqrt{A/(A+B)}$ is real, and
Eq. \ref{scalar field localised gravity} yields a kink-like
profile (Fig. \ref{localAdSkink}). This profile can be numerically
inverted to give a plot of the generating potential $\hat{V}(\hat{\Phi})$ as a
function of the scalar field $\hat{\Phi}$ (Fig. \ref{localAdSV}). As noted previously, 
the case $0 < A < B$ yields
a metric qualitatively similar to that of \cite{karch+randall}, which
is expected to localise gravity \cite{soda}.

\FIGURE{
\includegraphics[angle=270,width=0.6\textwidth]{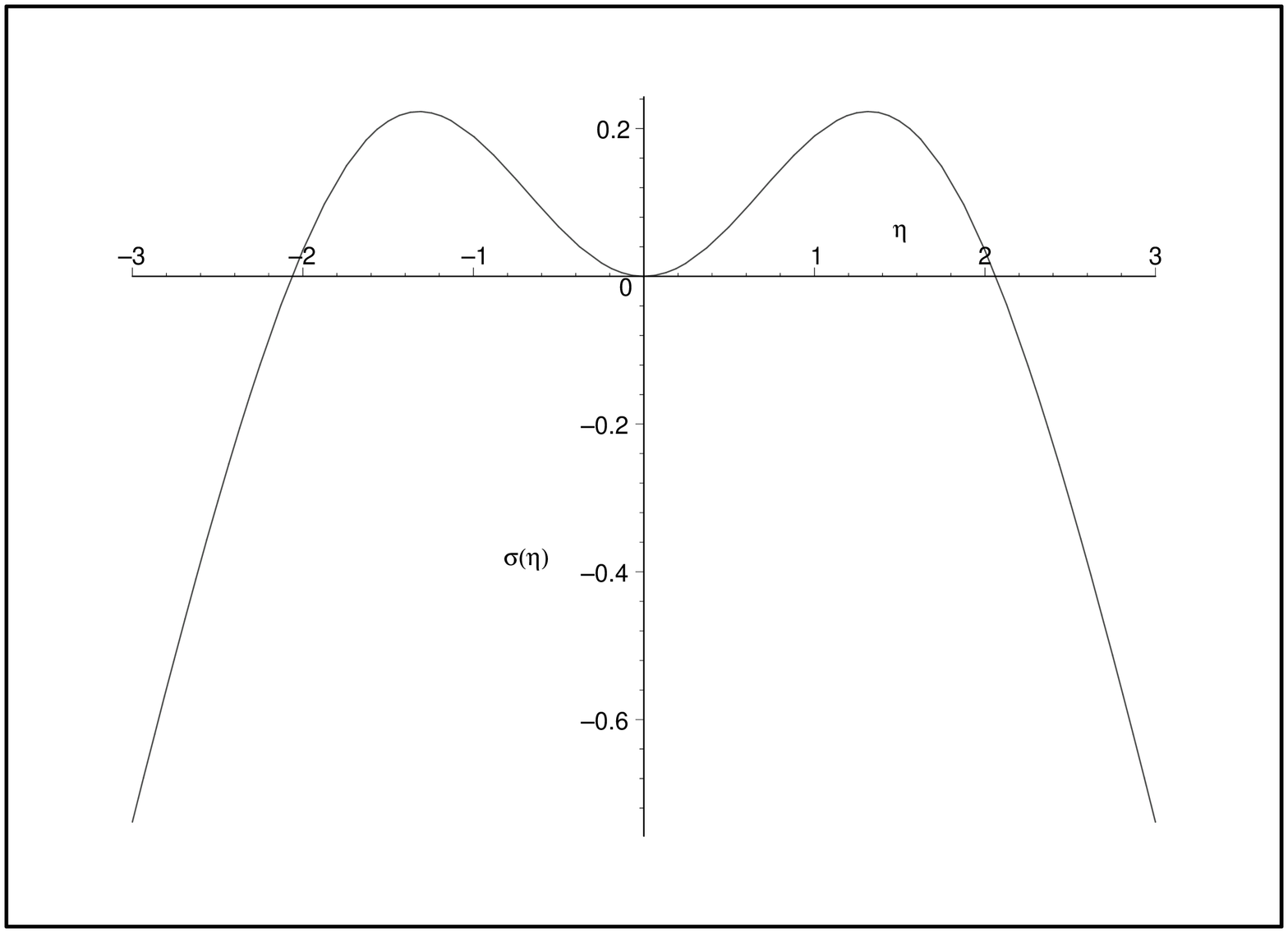}
\caption{Warp factor exponent $\sigma(\eta)$ for a domain wall with AdS$_4$ brane cosmology;
$A=0.2,B=0.8$, see Eq. \ref{locally localised gravity ansatz}.}
\label{localAdSmetric}}

\FIGURE{
\includegraphics[angle=270,width=0.6\textwidth]{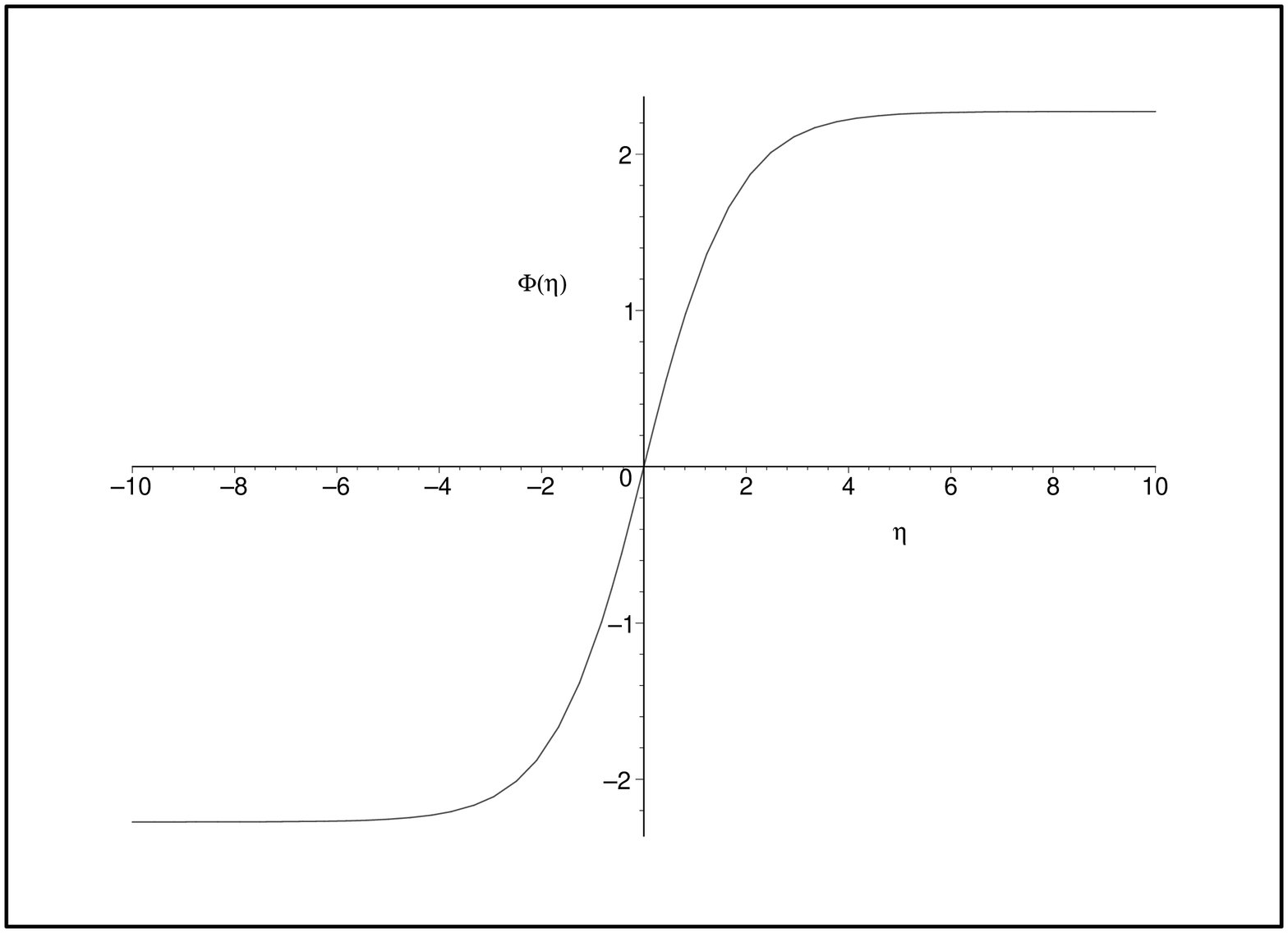}
\caption{Dimensionless scalar field profile $\hat{\Phi}(\eta)$ for a domain wall with AdS$_4$ brane cosmology;
$A=0.2,B=0.8$, see Eqs. \ref{locally localised gravity
ansatz}-\ref{scalar field localised gravity}.}
\label{localAdSkink}}

\FIGURE{
\includegraphics[angle=0,width=0.6\textwidth]{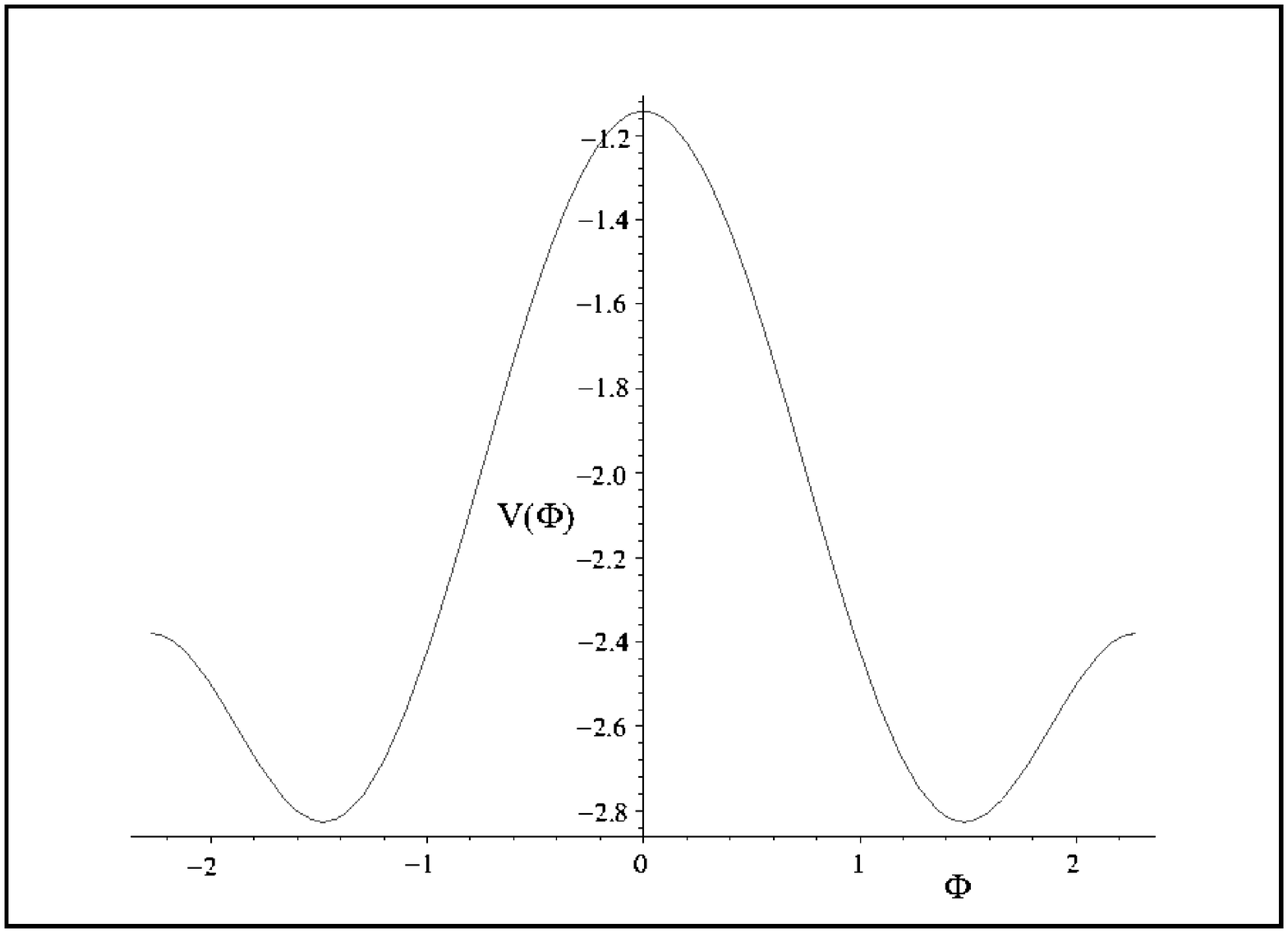}
\caption{$\hat{V}(\hat{\Phi})$ vs $\hat{\Phi}$ for a domain wall with AdS$_4$ brane cosmology;
$A=0.2,B=0.8$, see Eqs. \ref{locally localised gravity
ansatz}-\ref{scalar field localised gravity} .}
\label{localAdSV}}

In the special case where $B=0$, Eq. \ref{scalar field localised
gravity} becomes,
\begin{equation} \hat{\Phi}(\eta) = \pm 2 \arctan
\sinh (r \eta). \end{equation}
Then the potential $V(\Phi(y))$ can be obtained analytically from
Eq. \ref{locally localised V}, by noting that
\begin{equation} \textrm{sech}^2 (r \eta)  =  \frac{1}{1 + \sinh^2 (r
\eta)} = \cos^2 \left(\frac{\hat{\Phi}}{2} \right), \end{equation}
and so in this special case, the dimensionless potential takes the form,
\begin{equation} \hat{V}(\eta) =  - 2 r^2 \left[ 1 + 3
\cos^2 \left(\frac{\hat{\Phi}}{2} \right) \right]. \label{simple
analytic V} \end{equation}

In fact, it is possible to obtain a slightly more general solution in
this case: if we do not require Eq. \ref{condition on parameters} to
hold, then the scalar field configuration for $B=0, \gamma < 0$ becomes,
\begin{equation} \hat{\Phi}(\eta) = \frac{\pm \sqrt{1 - A^2 r^2}}{A r}
\arctan \sinh (r \eta), \end{equation}
and Eq. \ref{dimensionless equation for V} yields,
\begin{equation} \hat{V}(\hat{\Phi}) = - 2 r^2 \left[ 1 - \frac{3}{4} \left(
1 - \frac{1}{A^2 r^2} \right) \cos^2 \left( \frac{A R}{\sqrt{1 - A^2
r^2 }} \hat{\Phi} \right) \right], \end{equation}
which reduces to Eq. \ref{simple analytic V} in the case where $5 A^2
r^2 = 1$.  Note that in these cases the asymptotes of the scalar field do
not coincide with local minima of the potential.

Now in the dS case ($\gamma > 0$), it follows from Eq. \ref{dS consistency condition} that $4A (A + B) < 0$, and
consequently $A/(A+B) < 0$. A sample profile of the warp factor exponent is shown in
Fig. \ref{localdSmetric}: as expected from the results of
Sec. \ref{properties of solutions to the Einstein equations with
dS$_4$ cosmology}, it diverges to $\infty$ a finite distance from the
brane. (The warp factor exponent is obtained by taking $- \ln
\omega(\eta)$; we plot the exponent rather than the warp factor itself
to facilitate comparison with previous studies.)

\FIGURE{
\includegraphics[angle=270,width=0.6\textwidth]{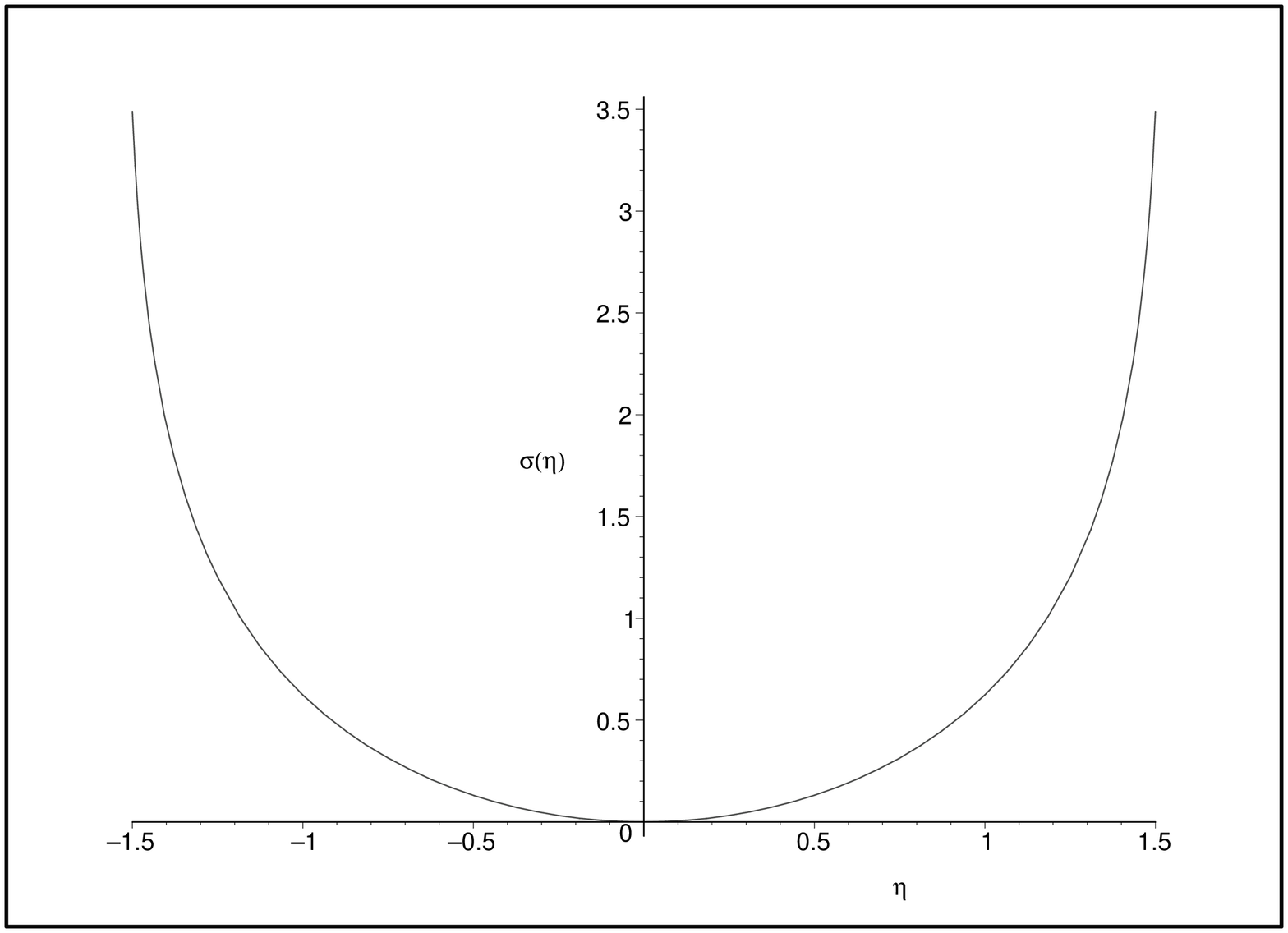}
\caption{Warp factor exponent $\sigma(\eta)$ for a brane with dS$_4$ cosmology lying between two naked singularities;
$A=-1,B=2$, see Eq. \ref{locally localised gravity ansatz}.}
\label{localdSmetric}}

Writing $i \alpha = \sqrt{A/(A+B)}$, for
$\alpha$ real, the
scalar field profile of Eq. \ref{scalar field localised gravity} becomes,
\begin{equation}  \hat{\Phi}(\eta) = \pm \left[ \arctan
\sinh (r \eta) - \alpha \, \textrm{arctanh} \left( \alpha
\sinh (r \eta) \right) \right]. \label{singular scalar field} \end{equation}
Let us assume a $+$ sign for the purpose of this analysis: there are
no significant differences between the kink and antikink solutions. A
sample profile is given in Fig. \ref{localdSkink}.

\FIGURE{
\includegraphics[angle=270,width=0.6\textwidth]{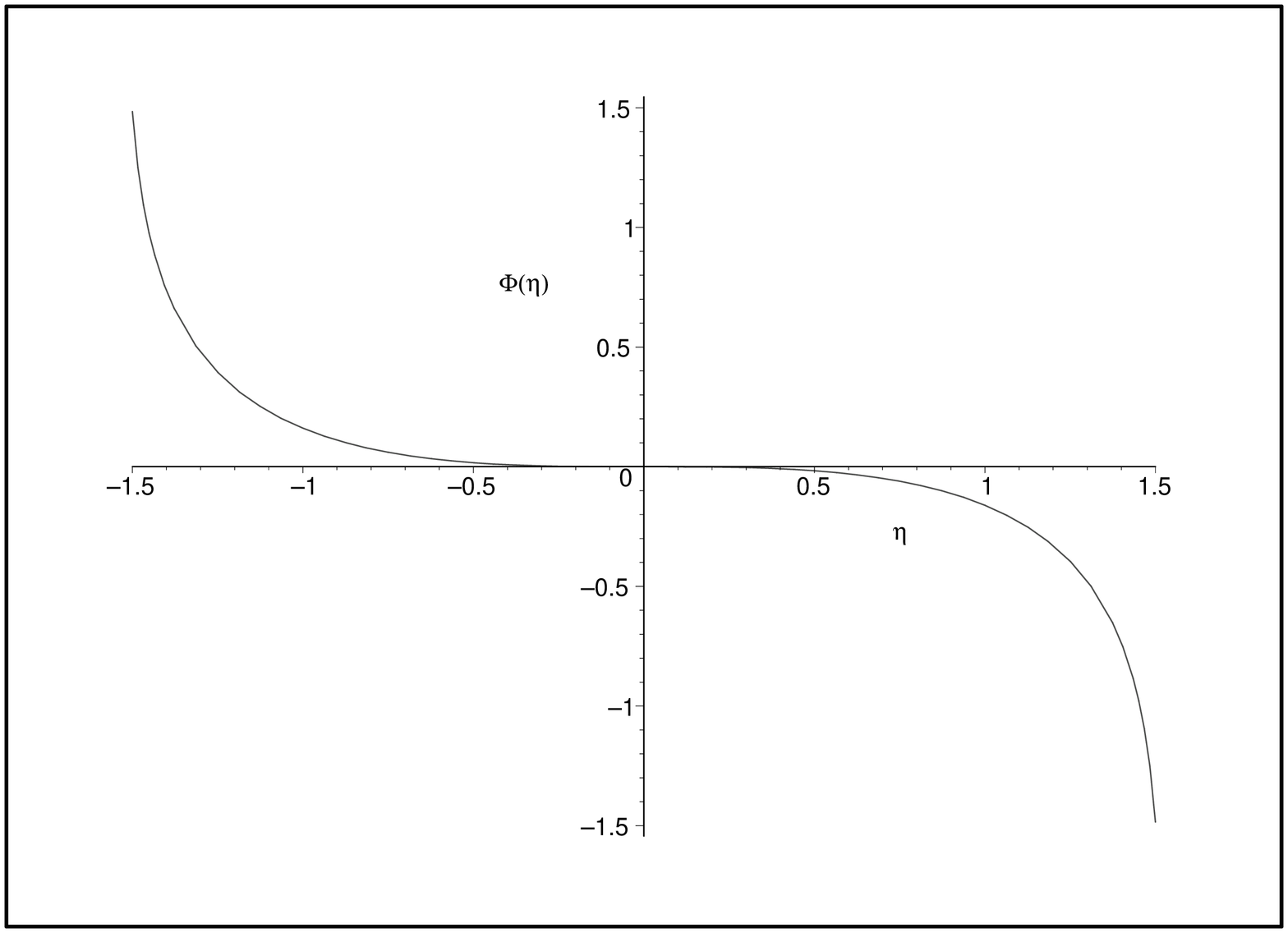}
\caption{Dimensionless scalar field profile $\hat{\Phi}(\eta)$ supporting a brane with dS$_4$ cosmology;
$A=-1,B=2$, see Eqs. \ref{locally localised gravity ansatz},
\ref{singular scalar field}.}
\label{localdSkink}}

As is clear from Fig. \ref{localdSkink}, the second term diverges to $\pm \infty$ where $\alpha \sinh (r \eta) =
\pm 1$: it is trivial to check that the points of divergence correspond
precisely with the zeroes of the metric. Close to these singular
points, which we shall denote $\pm \eta_0$, $\hat{\Phi}(\eta)$ can be approximated
by,
\begin{eqnarray} \hat{\Phi}(-\eta_0 + \Delta \eta) & = & -  \arctan \left(
\frac{1}{\alpha} \right) - \frac{\alpha}{2} \ln \left( \frac{1}{2} r
\alpha \sqrt{1+\alpha^2} \Delta \eta \right) + O(\Delta \eta),
\label{Phi near singularity 1} \\ \hat{\Phi}(\eta_0 -
\Delta \eta) & = & \arctan \left( \frac{1}{\alpha} \right) +
\frac{\alpha}{2} \ln \left(  \frac{1}{2} r
\alpha \sqrt{1+\alpha^2} \Delta \eta \right) +O(\Delta
\eta). \label{Phi near singularity 2} \end{eqnarray}
Inverting these expressions allows the potential $\hat{V}$ to be
written analytically as
a function of $\hat{\Phi}$, close to the singularities. We find that the
leading order term in $\hat{V}(\hat{\Phi})$ is given by,
\begin{equation} \hat{V}(\hat{\Phi}) \sim - \frac{3}{32} \alpha^4 \left(1 +
\alpha^2 \right) e^{- \frac{4}{\alpha} \arctan \left( \frac{1}{\alpha}
\right) } e^{\frac{4}{\alpha} |\hat{\Phi}| }, \end{equation}
so in this case $\hat{V}(\hat{\Phi})$ is unbounded below, and the scalar field
diverges a finite distance from the brane. The potential can also be
plotted numerically against the scalar field over the range between
the singularities (Fig. \ref{localdSV}), but there are no other
features of real interest.

\FIGURE{
\includegraphics[angle=0,width=0.6\textwidth]{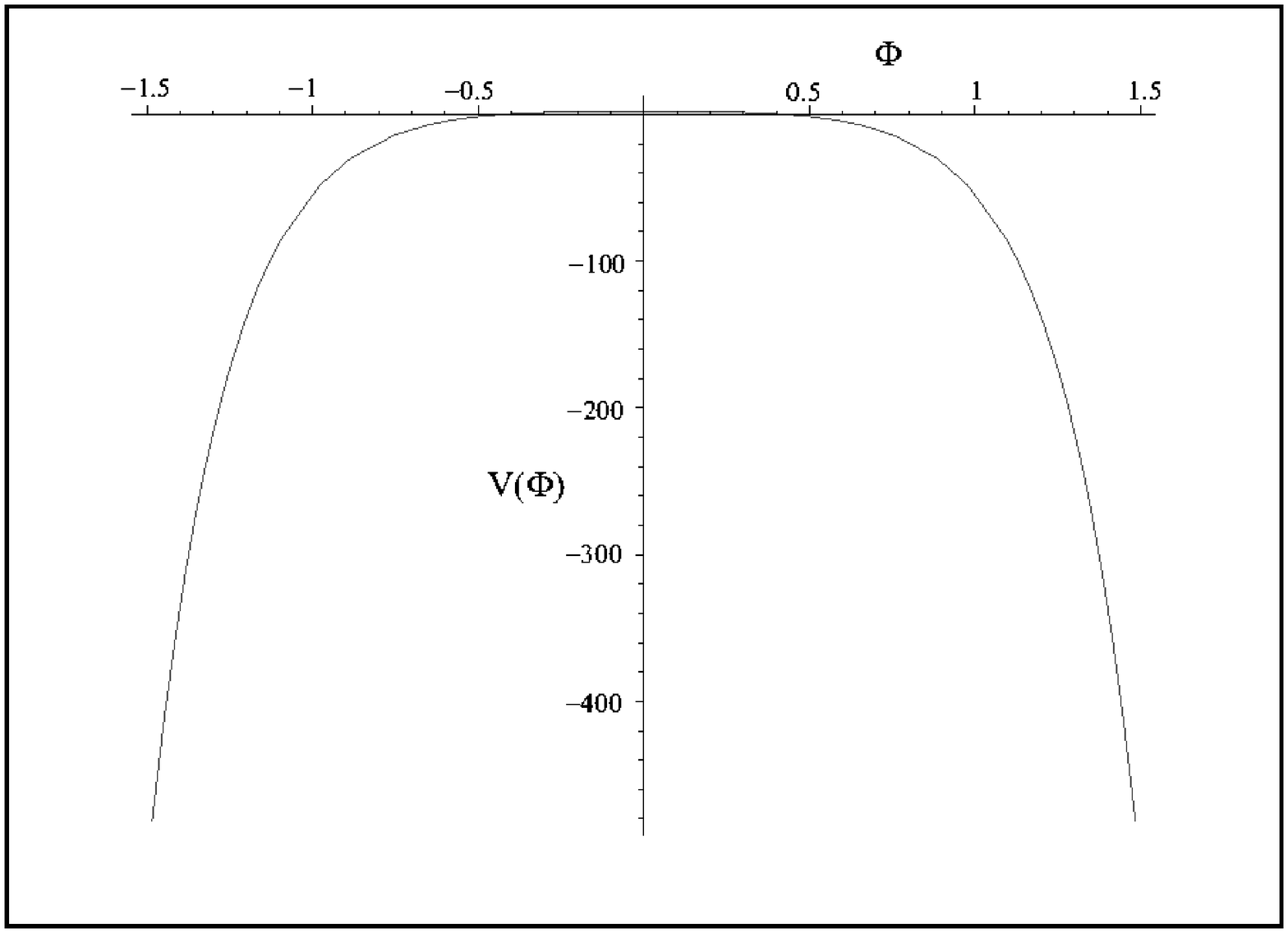}
\caption{$\hat{V}(\hat{\Phi})$ vs $\hat{\Phi}$ for a potential
supporting a brane with dS$_4$ cosmology;
$A=-1,B=2$, see Eq. \ref{locally localised gravity ansatz},
\ref{singular scalar field}.}
\label{localdSV}}

The 5D curvature scalar is given by Eq. \ref{5D curvature scalar -
dimensionless}, and close to the singularities the leading order term is of the form,
\begin{equation} R^{(5)}(\eta) \sim  |\gamma| \alpha^2
\frac{1}{\left( \Delta \eta \right)^2}. \end{equation}
Thus in this case the zeroes in the metric represent curvature
singularities, and indicate divergences in the scalar field
energy-momentum tensor, rather than merely being horizons.

\section{Fermion trapping by warped metrics}

\label{fermion trapping in warped metrics}

The trapping of fermions on a scalar field domain wall was first
described in the 1980s (\cite{rubakov+shaposhnikov, jackiw+rebbi}). More recently, fermion
trapping in a scalar field domain wall with a five-dimensional warped
metric has been discussed by a number of authors (\cite{ringeval,
randjbar-daemi+shaposhnikov, kehagias+tamvakis, koley+kar1,
 bajc+gabadadze}), but to our knowledge only in the case of a
time-independent metric.

To write down the Dirac Lagrangian in curved spacetime, we employ the
vielbeins $V^{A}_{M}$ and inverse vielbeins $V_{A}^{M}$ defined by (\cite{weinberg, birrell}):
\begin{equation} g_{M N}(x) = V^{A}_{M}(x)
V^{B}_{N}(x) \eta_{A B}. \end{equation}
Here $\eta_{A B}$ is the Minkowski space metric, and $A, B, C, ...$ indicate coordinates in (locally defined) 5D Minkowski space, with $M, N, ...$
indicating coordinates in the curved space described by the metric
$g_{M N}$.

The Dirac Lagrangian can then be written out as,
\begin{equation} \mathcal{L}_{\Psi} \equiv \bar{\Psi} \Gamma^{A}
\mathcal{D}_{A} \Psi - g_F \bar{\Psi} F(\Phi) \Psi, \end{equation}
where the $\Gamma$'s are the Minkowski space Dirac matrices,  $F$ is some odd function of $\Phi$ describing the coupling between
the Dirac field and the scalar field, and
$\mathcal{D}_{A}$ is the covariant derivative with spin connection defined by,
\begin{eqnarray} \mathcal{D}_{A} & \equiv & V_{A}^{M} \left(
\frac{\partial}{\partial x^{M}} + \Sigma_{M}(x) \right), \\
\Sigma_{M}(x) & \equiv & \frac{1}{2} \sigma^{B C}
V_{B}^{N} V_{C N ; M}, \\  V_{C N ; M} & = & \frac{\partial V_{C N}}{\partial x^{M}} -
\Gamma^{R}_{N M} V_{C R}, \\ \Gamma^{R}_{N M} & = & \frac{1}{2} g^{S R} \left[ \frac{\partial g_{M
N}}{\partial x^{N}} +  \frac{\partial g_{N S}}{\partial x^{M}} -
\frac{\partial g_{M N}}{\partial x^{S}} \right]. \end{eqnarray}
Here $\sigma^{A B}$ describes how the field transforms under
infinitesimal Lorentz transformations, i.e. the spin of the field. For
a spin-$1/2$ field, $\sigma^{A B} = (1/4) [ \Gamma^{A},
\Gamma^{B} ]$.

We may choose the Dirac matrices in
five-dimensional Minkowski space, $\Gamma^{A}$, to be:
\begin{equation} \Gamma^{\alpha}  =  \gamma^{\alpha}, \qquad \Gamma^{4}  =  \gamma_5. \end{equation}
Here the $\gamma$ matrices are simply the 4-dimensional Dirac
matrices, which obey the four-dimensional Clifford algebra $\{
\gamma^{\alpha}, \gamma^{\beta} \} = 2 \eta^{\alpha \beta}$ with our
chosen metric signature $(- + + +)$.

It is easily verified that these $\Gamma^{A}$ matrices obey the five-dimensional Clifford algebra,
$\{ \Gamma^{A}, \Gamma^{B} \} = 2 \eta^{A B}$, and it
follows that for a Dirac field, if $A \ne B \Rightarrow \sigma^{A B} = (1/2)
\Gamma^{A} \Gamma^{B}$ (for $A = B$, of course,
$\sigma^{A B} = 0$). Note also that raising or lowering the
indices on the $\Gamma$'s is trivial: $\Gamma^0 = - \Gamma_0$,
$\Gamma^{A} = \Gamma_{A}$ for $A \ne 0$.

As previously, we consider a 5D metric of the form given in Eq.
\ref{general warped metric}. For now, we shall employ the $\sigma(y)$
parameterisation, giving the key equations rewritten in terms of $\omega(y)$
at the end of this section.
Let us choose a set of 4D vielbeins
$v^{\alpha}_{\mu}$ satisfying,
\begin{equation} g^{(4)}_{\mu \nu} = v^{\alpha}_{\mu} v^{\beta}_{\nu}
\eta_{\alpha \beta}. \end{equation}
Then the 5D vielbeins and inverse vielbeins may be chosen as,
\begin{equation} V^{A}_{M} =  \pmatrix{ e^{- \sigma(y)} v^{\alpha}_{\mu} & 0  \cr
0 & 1 }, \qquad  V_{A}^{M} =  \pmatrix{ e^{\sigma(y)} v_{\alpha}^{\mu} & 0  \cr
0 & 1 } . \label{5d vielbeins} \end{equation}

With this choice of vielbeins, the five-dimensional spin connection $\Sigma_{M}$
becomes:
\begin{equation} \Sigma_{M} = \pmatrix{  \Sigma_{\mu} + \frac{1}{2}
\sigma'(y) e^{- \sigma(y)} \Gamma^{4} \Gamma^{\alpha}
v_{\alpha \mu} \cr 0}, \label{spin connection} \end{equation}
where $\Sigma_{\mu}$ is the spin connection for fermion fields
in the four-dimensional spacetime described by the metric
$g^{(4)}_{\mu \nu}$.

Consequently, the covariant derivative with spin connection becomes
\begin{equation} \mathcal{D}_{\alpha} =  e^{\sigma(y)}
v_{\alpha}^{\mu} \left(\frac{\partial}{\partial x^{\mu}} +   \Sigma_{\mu} + \frac{1}{2}
\sigma'(y) e^{- \sigma(y)} \Gamma^{4} \Gamma^{\beta}
v_{\beta \mu} \right), \qquad \mathcal{D}_4  = 
\frac{\partial}{\partial y}, \end{equation} 
and the sum $\Gamma^{A} \mathcal{D}_{A}$ can be written,
\begin{equation} \Gamma^{A} \mathcal{D}_{A}  =  e^{\sigma(y)} \left[ \gamma^{\alpha} v_{\alpha}^{\mu} \left(\frac{\partial}{\partial
x^{\mu}} +  \Sigma_{\mu} \right) \right] + \gamma_5
\left(\frac{\partial}{\partial y} - 2 \sigma'(y) \right).
\label{slashed covariant derivative with spin connection} \end{equation} 

The five-dimensional Dirac equation for a fermion field coupled to
gravity and the background scalar field $\Phi$ is simply,
\begin{equation}  \left( \Gamma^{A}
\mathcal{D}_{A} - g_F F(\Phi) \right) \Psi = 0. \label{Dirac equation}
\end{equation}

It is well known (\cite{ringeval, randjbar-daemi+shaposhnikov, koley+kar1, bajc+gabadadze}) that the confinement of fermion modes to the brane
depends on their chirality. Let us therefore write the Dirac spinor $\Psi$ in
the form
\begin{equation} \Psi(x) = \mathcal{U}_L(y) \psi_L(t,x^{i}) +
\mathcal{U}_R(y) \psi_R(t,x^{i}), \label{separable Dirac field ansatz} \end{equation}
where $\psi_L$ and $\psi_R$ are the left-handed and right-handed
components of a four-dimensional Dirac field, and hence $\gamma_5
\psi_L = - \psi_L$, $\gamma_5 \psi_R = \psi_R$. Inserting this ansatz
and Eq. \ref{slashed covariant derivative with spin connection}
into the Dirac equation yields,
\begin{eqnarray} 0 & = &  \left( \Gamma^{A}
\mathcal{D}_{A} - g_F F(\Phi) \right) \Psi \nonumber \\ & = & 
\mathcal{U}_L(y) e^{\sigma(y)} \left[ \gamma^{\alpha} v_{\alpha}^{\mu} \left(\frac{\partial}{\partial
x^{\mu}} +   \Sigma_{\mu} \right) \right] \psi_L(t,x^{i}) - \nonumber \\ &
& -  \psi_L(t,x^{i})
\left(\frac{\mathrm{d}}{\mathrm{d} y} - 2 \sigma'(y) \right)
\mathcal{U}_L(y) + \nonumber \\ & & +  
\mathcal{U}_R(y) e^{\sigma(y)} \left[ \gamma^{\alpha} v_{\alpha}^{\mu} \left(\frac{\partial}{\partial
x^{\mu}} +   \Sigma_{\mu} \right) \right] \psi_R(t,x^{i}) +
\nonumber  \\ &
& + \psi_R(t,x^{i})
\left(\frac{\mathrm{d}}{\mathrm{d} y} - 2 \sigma'(y) \right)
\mathcal{U}_R(y) - \nonumber \\ & & - g_F F(\Phi) \mathcal{U}_L(y)
\psi_L(t,x^{i}) - g_F F(\Phi)
\mathcal{U}_R(y) \psi_R(t,x^{i}). \end{eqnarray}

Now suppose we require that the four-dimensional spinors $\psi_L$,
$\psi_R$ satisfy the Dirac equation for fermions in the
four-dimensional spacetime described by the metric $g^{(4)}_{\mu \nu}$,
\begin{eqnarray}   \gamma^{\alpha} v_{\alpha}^{\mu}
\left( \partial_{\mu} + \Sigma_{\mu} \right) \psi_L & = & m
\psi_R, \\  \gamma^{\alpha} v_{\alpha}^{\mu}
\left( \partial_{\mu} + \Sigma_{\mu} \right) \psi_R & = & m
\psi_L. \end{eqnarray}

Then the five-dimensional Dirac equation becomes,
\begin{eqnarray} 0 & = &  \psi_R(t,x^i) \left[ m e^{\sigma(y)}
\mathcal{U}_L(y) + \left(\frac{\mathrm{d}}{\mathrm{d} y} - 2 \sigma'(y) \right)
\mathcal{U}_R(y) -  g_F F(\Phi)
\mathcal{U}_R(y) \right] +  \nonumber \\ &
& + \psi_L(t,x^i) \left[  m e^{\sigma(y)}
\mathcal{U}_R(y)  - \left(\frac{\mathrm{d}}{\mathrm{d} y} - 2 \sigma'(y) \right)
\mathcal{U}_L(y) - g_F F(\Phi) \mathcal{U}_L(y) \right], \end{eqnarray}
and equating the coefficients of $\psi_R$ and $\psi_L$ separately to
zero, we obtain a pair of coupled first-order differential equations
in the bulk coordinate,
\begin{eqnarray}  m e^{\sigma(y)}
\mathcal{U}_L(y) + \left(\frac{\mathrm{d}}{\mathrm{d} y} - 2 \sigma'(y) \right)
\mathcal{U}_R(y) -  g_F F(\Phi)
\mathcal{U}_R(y) & = & 0 \, , \label{massive modes 1} \\  m e^{\sigma(y)}
\mathcal{U}_R(y)  -  \left(\frac{\mathrm{d}}{\mathrm{d} y} - 2 \sigma'(y) \right)
\mathcal{U}_L(y) - g_F F(\Phi) \mathcal{U}_L(y) & = &
0 \, . \label{massive modes 2} \end{eqnarray}
Note that these equations are identical to those for a static domain
wall (see for example \cite{ringeval}): the generalisation to any 4D metric
$g^{(4)}_{\mu \nu}$ only modifies the warp factor $\sigma(y)$ and the scalar
field $\Phi$.

Let us now define the dimensionless rescaled mass and coupling
constant for the case $\gamma \ne 0$, in the case where $F(\Phi) = \Phi$,
\begin{equation} \mu = m \sqrt{\frac{3}{|\gamma|}}, \qquad h_F =
\frac{3 g_F}{\kappa \sqrt{|\gamma|}}. \label{h_F and mu} \end{equation}

Then Eq. \ref{massive modes 1}-\ref{massive modes 2} can be rewritten
in terms of dimensionless quantities,
\begin{eqnarray}  \mu e^{\sigma(\eta)}
\mathcal{U}_L(\eta) +  \left(\frac{\mathrm{d}}{\mathrm{d} \eta} - 2 \sigma'(\eta) \right)
\mathcal{U}_R(\eta) -   h_F  \hat{\Phi}(\eta)
\mathcal{U}_R(\eta) & = & 0, \label{dimensionless massive modes 1} \\  \mu e^{\sigma(\eta)}
\mathcal{U}_R(\eta)  -  \left(\frac{\mathrm{d}}{\mathrm{d} \eta} - 2 \sigma'(\eta) \right)
\mathcal{U}_L(\eta) - h_F \hat{\Phi}(\eta) \mathcal{U}_L(\eta) & = &
0. \label{dimensionless massive modes 2} \end{eqnarray}

As previously, in configurations containing bulk singularities we may
wish to rewrite the metric in terms of the warp factor $\omega$: the only reason
for casting our equations in terms of the warp factor
exponent is for easy comparison with the existing literature. In
this case the mode equations can be derived as,
\begin{eqnarray} \frac{m}{\omega(y)} \mathcal{U}_L(y) +
\left(\frac{\mathrm{d}}{\mathrm{d} y} + 2 \frac{\omega'(y)}{\omega(y)} \right)
\mathcal{U}_R(y) -  g_F F(\Phi)
\mathcal{U}_R(y) & = & 0 \, , \label{massive modes 1 in omega} \\   \frac{m}{\omega(y)}
\mathcal{U}_R(y)  -  \left(\frac{\mathrm{d}}{\mathrm{d} y} + 2 \frac{\omega'(y)}{\omega(y)} \right)
\mathcal{U}_L(y) - g_F F(\Phi) \mathcal{U}_L(y) & = &
0 \, , \label{massive modes 2 in omega} \end{eqnarray}
or in terms of dimensionless quantities for the $F(\Phi)=\Phi$ case,
 \begin{eqnarray}  \frac{\mu}{\omega(\eta)}
\mathcal{U}_L(\eta) +  \left(\frac{\mathrm{d}}{\mathrm{d} \eta}  + 2 \frac{\omega'(\eta)}{\omega(\eta)} \right)
\mathcal{U}_R(\eta) -   h_F  \hat{\Phi}(\eta)
\mathcal{U}_R(\eta) & = & 0, \label{dimensionless massive modes 1 in omega} \\  \frac{\mu}{\omega(\eta)}
\mathcal{U}_R(\eta)  -  \left(\frac{\mathrm{d}}{\mathrm{d} \eta}  + 2
\frac{\omega'(\eta)}{\omega(\eta)} \right)
\mathcal{U}_L(\eta) - h_F \hat{\Phi}(\eta) \mathcal{U}_L(\eta) & = &
0. \label{dimensionless massive modes 2 in omega} \end{eqnarray}

\section{Confinement of the fermion zero mode on AdS$_4$ and dS$_4$ branes}

\label{confinement of fermion zero mode}

In the case where $m = 0$, the equations for $\mathcal{U}_L$ and
$\mathcal{U}_R$ simplify by decoupling,
\begin{eqnarray}  
\frac{\mathrm{d} \mathcal{U}_L(y)}{\mathrm{d} y} & = &  \mathcal{U}_L(y) \left(
2 \sigma'(y)  -  g_F F(\Phi(y)) \right),   \\ 
\frac{\mathrm{d} \mathcal{U}_R(y)}{\mathrm{d} y} & = &
\mathcal{U}_R(y) \left(2 \sigma'(y)  + g_F F(\Phi(y)) \right), \end{eqnarray}
allowing us to analytically study the confinement of the fermion zero mode.  We are
primarily interested in the behaviour of the fermion field within the region
close to the brane and, in the case of dS brane cosmology, bounded by
the singularities.  We shall thus use the $\sigma$ parameterisation, and demonstrate
that the presence of even the coordinate singularities discussed in Sec.\ 
\ref{Partially analytic solution with dS$_4$ slices and no curvature singularities}
leads to the failure of the usual confinement mechanism.

The above first-order linear differential equations can easily be solved
for the bulk coefficient functions,
\begin{equation} \mathcal{U}_L(y)  = A_L e^{2 \sigma(y)} e^{- g_F
\int F(\Phi(y))
\mathrm{d}y}, \qquad
\mathcal{U}_R(y)  = A_R e^{2 \sigma(y)} e^{ g_F
\int F(\Phi(y))
\mathrm{d}y}. \end{equation}

If we substitute these expressions back into the kinetic-energy
term of the Dirac Lagrangian, we obtain, 
\begin{eqnarray} \bar{\Psi} \Gamma^{A}
\mathcal{D}_{A} \Psi & = & A_L^2 e^{5 \sigma(y)} e^{- 2 g_F \int
F(\Phi(y)) \mathrm{d} y}  \bar{\psi}_L(t,x^{i}) \left[
\gamma^{\alpha} v_{\alpha}^{\mu} \left(\frac{\partial}{\partial
x^{\mu}} +   \Sigma_{\mu} \right) \right] \psi_L(t,x^{i}) +  \nonumber \\ &
&  +  A_R^2 e^{5 \sigma(y)} e^{2 g_F \int
F(\Phi(y)) \mathrm{d} y}  \bar{\psi}_R(t,x^{i}) \left[ 
\gamma^{\alpha} v_{\alpha}^{\mu} \left(\frac{\partial}{\partial
x^{\mu}} +   \Sigma_{\mu} \right) \right] \psi_R(t,x^{i}) +
 \nonumber \\  & &   + g_F
F(\Phi) A_L A_R e^{4 \sigma(y) } \left( \bar{\psi}_R(t,x^i) \psi_L(t,x^i) +  \bar{\psi}_L(t,x^i) \psi_R(t,x^{i}) \right).  \end{eqnarray}

The action for the fermion field is simply,
\begin{equation} S_{\Psi} = \int \mathrm{d}^5 x \sqrt{-g(x)}
\mathcal{L}_{\Psi}(x). \end{equation}
If $\hat{g} = Det(g^{(4)}_{\mu \nu})$, then $g = e^{- 8 \sigma(y)}
\hat{g}$, and $\sqrt{-g} = e^{-4 \sigma(y)} \sqrt{- \hat{g}}$.  Noting that the term in $F(\Phi(y))$ is odd for a kink solution, we need
only consider the first two terms in the action, because the third
integrates to zero. These terms factorise
into the usual 4D action over
the brane coordinates,
\begin{equation}  \int^\infty_{- \infty} \mathrm{d}^4 x \sqrt{-
\hat{g}}  \left[ \bar{\psi}_{L/R}(t,x^{i}) \left( 
\gamma^{\alpha} v_{\alpha}^{\mu} \left(\frac{\partial}{\partial
x^{\mu}} +   \Sigma_{\mu} \right) \right) \psi_{L/R}(t,x^{i}) \right],  \end{equation}
multiplied by  integrals over the bulk coordinates given by
\begin{equation} A_L^2 \int^\infty_{- \infty} \mathrm{d}y \, \, e^{ \sigma(y)} e^{ - 2 g_F
\int F(\Phi(y)) \mathrm{d}y}, \qquad  A_R^2 \int^\infty_{- \infty}
\mathrm{d}y \, \, e^{ \sigma(y)} e^{2 g_F
\int F(\Phi(y)) \mathrm{d} y}, \label{normalisation integral} \end{equation}
for left-handed and right-handed fermion fields respectively.

For the case where $F(\Phi) = \Phi$ and $\gamma \ne 0$, the
normalisation integrals can be written in terms of the dimensionless
quantities,
\begin{equation} A_L^2 \sqrt{\frac{3}{|\gamma|}}
\int^\infty_{- \infty} \mathrm{d}\eta \, \, e^{ \sigma(\eta)} e^{ - 2 h_F
\int \hat{\Phi}(\eta) \mathrm{d} \eta}, \qquad  
A_R^2 \sqrt{\frac{3}{|\gamma|}} \int^\infty_{- \infty}
\mathrm{d}\eta \, \, e^{ \sigma(\eta)} e^{2 h_F
\int \hat{\Phi}(\eta) \mathrm{d} \eta}. \label{dimensionless normalisation integral} \end{equation}

It is now clear that in general,
zeroes in the metric (points where $\sigma(\eta)$ diverges to $\infty$)
correspond to singularities in the integrand, and thus the
kinetic energy part of the action may be expected to be non-normalisable in
cases where the metric contains zeroes. In particular, this will be the case
for the new solution presented in Eqs. \ref{bruteforce-metric-ansatz}-\ref{sampleparameters}.
This obviously represents a challenge for that kind of model: although it has
the nice properties that the bulk inverse-metric singularity is a coordinate not a curvature singularity
and the scalar field configuration is smooth, it has the
drawback that the usual kink-based fermion zero-mode localisation mechanism does not generalise
to that dS$_4$ case.  Dealing with this challenge is beyond the scope of this paper,
but two logical approaches immediately suggest themselves:  One could look to
introduce new physics, beyond the scalar kink, to localise fermions.
Alternatively, we know that our universe is only approximately de Sitter, so it
would be interesting to explore fermion localisation via the kink mechanism
in an effective domain wall FRW cosmology that has the transitions from
radiation to matter to vacuum energy domination of regular cosmology.

It seems possible that if the
kink $\Phi$ also diverges to $\pm \infty$ at the metric zeroes, then one of the
chiral fermion zero modes might be normalisable over the bulk.
However, in
the dS$_4$ solution we have described with curvature singularities at
the metric zeroes, this cancellation does not occur. Eqs. \ref{Phi
near singularity 1}-\ref{Phi near singularity 2} show that close to the
singularities, the leading order term of $\hat{\Phi}(\eta)$ is of the form $\ln
(q \Delta \eta)$, so the integral function $\int \hat{\Phi}(\eta) \,
\mathrm{d} \eta$ is bounded close to the singularities, while
$e^{\sigma(\eta)}$ diverges as $1 / \Delta \eta$.

If the warp factor exponent $\sigma(y)$ is smooth and nonsingular, the fermion
zero modes may or may not be normalisable. Suppose $\sigma(y) \sim c |y|$ as $y \rightarrow \pm
\infty$, for some $c < 0$. Then one of the two chiral fermion zero modes will always be confined:
the chirality of this mode will depend on the scalar field profile. The
other mode may or may not be confined, depending on the asymptotic behaviour of
$\int \mathrm{d} y \, F(\Phi(y))$ relative to $\sigma(y)$. This is the case
when the metric diverges to infinity asymptotically, and is relevant
for the cases considered in Sec. \ref{locally localised gravity}, with AdS$_4$ cosmology on the
brane. This behaviour has also been studied by Koley and Kar in the
context of a ``ghost'' scalar field with negative energy in the bulk
in \cite{koley+kar2}, and by Bajc and Gabadadze for fermions localised
on a non-fine-tuned RS brane by a scalar field in \cite{bajc+gabadadze}.

If conversely the
warp factor exponent behaves as $\sigma(y) \sim c |y|$ for $y \rightarrow \pm
\infty$, with $c > 0$, then one of the two modes will certainly be
non-normalisable, while the other may be normalisable depending on the
behaviour of the integral  $\int \mathrm{d} y \, F(\Phi(y))$. This is the
case for fine-tuned static brane solutions
(\cite{ringeval, jackiw+rebbi, bajc+gabadadze}), where the metric elements approach zero for large
$|y|$. This type of warp factor is employed to localise gravity on the
brane in the Randall-Sundrum approach (\cite{RS2}).

In particular, if $F(\Phi(y))$ has a kink or anti-kink
profile, and $\lim_{y \rightarrow \infty} F(\Phi(y)) = - \lim_{y \rightarrow
-\infty} F(\Phi(y)) = \lambda$, then far from the brane, $\int \mathrm{d} y
F(\Phi(y)) \sim \lambda |y|$ (up to a constant of integration). Consequently,
if $\sigma(y) \sim c |y| $ for $|y|$ large, then the normalisation integrands are of the form,
\begin{equation} \exp \left( \left( c + 2 g_F \lambda \right) |y|
\right), \qquad \exp \left( \left( c - 2 g_F \lambda \right) |y| \right), \end{equation}
for right- and left-handed fields respectively.

For the ``locally localised gravity'' AdS$_4$ solution outlined in
Sec. \ref{locally localised gravity}, these conditions hold for the simple
coupling $F(\Phi) = \Phi$, with the parameters,
\begin{equation} \lambda = \frac{\sqrt{3}}{\kappa} \frac{\pi}{2} \left( 1
+ \alpha \right), \qquad c = - r. \end{equation}
Thus the action is normalisable for right-handed fermion fields provided,
\begin{equation} g_F < \frac{\kappa}{\sqrt{3}} \frac{r}{\pi \left(1 +
\alpha \right)}, \end{equation}
and for left-handed fermion fields if
\begin{equation} g_F > - \frac{\kappa}{\sqrt{3}} \frac{r}{\pi \left(1 +
\alpha \right)}, \end{equation}
but the latter relation is always true for positive $g_F$.

In terms of the dimensionless quantities, we replace $y$ with $\eta$
and $g_F$ with $h_F$ in the discussion above, and the parameters become $c = - r$, $\lambda
= \pi / 2 (1 + \alpha)$. The normalisability conditions become,
\begin{equation} h_F <  \frac{r}{\pi \left(1 +
\alpha \right)} \end{equation}
for right-handed fermions, and
\begin{equation} h_F > - \frac{r}{\pi \left(1 +
\alpha \right)}, \end{equation}
for left-handed fermions.

However, normalisability of the chiral fermion zero-modes does not
necessarily imply that those modes are localised on the brane. For
example, in some parameter regimes,  the normalisation integrand has two peaks at the points where the warp factor
turns around, and a local minimum (rather than a local maximum) on the
brane, before falling off rapidly far from the brane. If the local
minimum is sufficiently shallow and the peaks are sufficiently close
to the brane, this behaviour may still constitute localisation to the brane.

The Lagrangian density always has a local extremum at the brane
(i.e. its derivative with respect to the bulk coordinate is zero
there): its second derivative with respect to the bulk coordinate
determines whether the brane lies at a local minimum or maximum. The
second derivative of the integrand of Eq. \ref{dimensionless normalisation integral}
has the same sign as
\begin{equation} \sigma''(\eta) \pm 2 h_F \hat{\Phi}'(\eta), \end{equation}
where as previously, the $+$ sign applies to right-handed fermion fields and the $-$
sign corresponds to left-handed fields.

For the AdS$_4$ case outlined in Sec. \ref{locally localised gravity},
\begin{equation}   \sigma''(0) \pm 2 h_F \hat{\Phi}'(0) = r^2
\left( 1
- 2 \alpha^2 \right) \pm 2 h_F r \left( 1 + \alpha^2 \right), \end{equation}
so the brane lies at a local maximum of the Lagrangian density for,
\begin{equation} \pm h_F < \frac{r \left( 2 \alpha^2 - 1 \right)}{2
\left( \alpha^2
+ 1 \right)}. \end{equation}
Note that in the case where $0 < A < B$, $\alpha < 1/2$ and therefore
the RHS of this expression is always negative. Consequently, this
condition cannot hold for right-handed fermion modes: this result is
analogous to the usual behaviour of fermion zero modes for a Minkowski
brane, where right-handed modes are always unconfined
(\cite{ringeval, bajc+gabadadze}). Fig.
\ref{rightfermionconfined} demonstrates a sample profile of the
integrand of Eq. \ref{dimensionless normalisation integral} for the
case where the right-handed fermion modes are normalisable, but not
confined to the brane. Left-handed fermion modes are always
normalisable, and may (Fig. \ref{leftfermionconfined}) or may not
(Fig. \ref{leftfermionunconfined}) exhibit a peak at the brane itself.

\FIGURE{
\includegraphics[angle=0,width=0.6\textwidth]{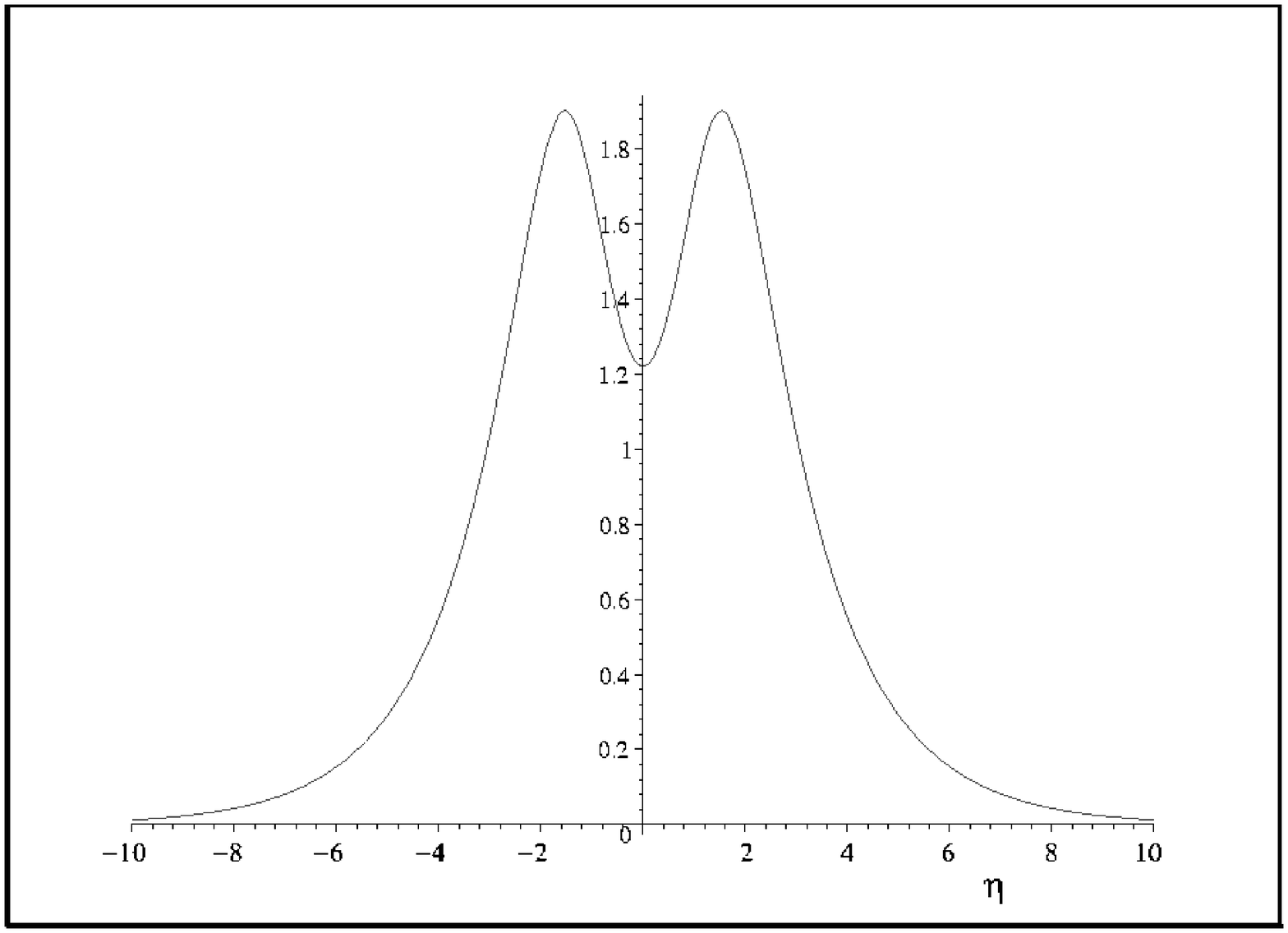}
\caption{Bulk dependence of the kinetic energy term, 
Eq. \ref{dimensionless normalisation integral}, for right-handed
fermions on a domain wall with AdS$_4$ brane cosmology;
$A=0.2,B=0.8,h_F=0.1$; see Eqs. \ref{locally localised gravity ansatz}-\ref{locally localised V} 
and Eq. \ref{h_F and mu}.}
\label{rightfermionconfined}}

\FIGURE{
\includegraphics[angle=0,width=0.6\textwidth]{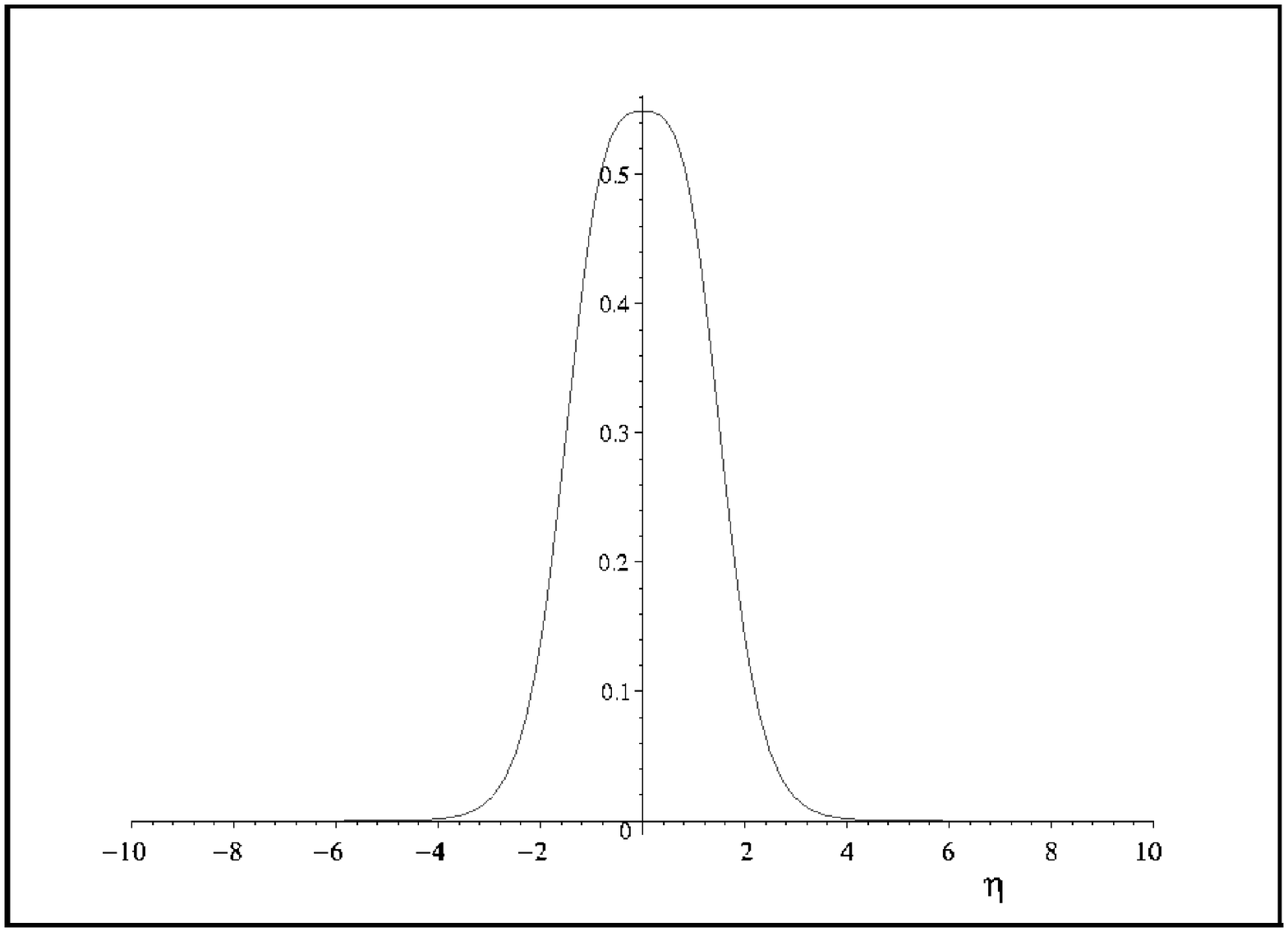}
\caption{Bulk dependence of the kinetic energy term, 
Eq. \ref{dimensionless normalisation integral}, for left-handed
fermions on a domain wall with AdS$_4$ brane cosmology;
$A=0.2,B=0.8,h_F=0.3$; see Eqs. \ref{locally localised gravity ansatz}-\ref{locally localised V} 
and Eq. \ref{h_F and mu}.}
\label{leftfermionconfined}}

\FIGURE{
\includegraphics[angle=0,width=0.6\textwidth]{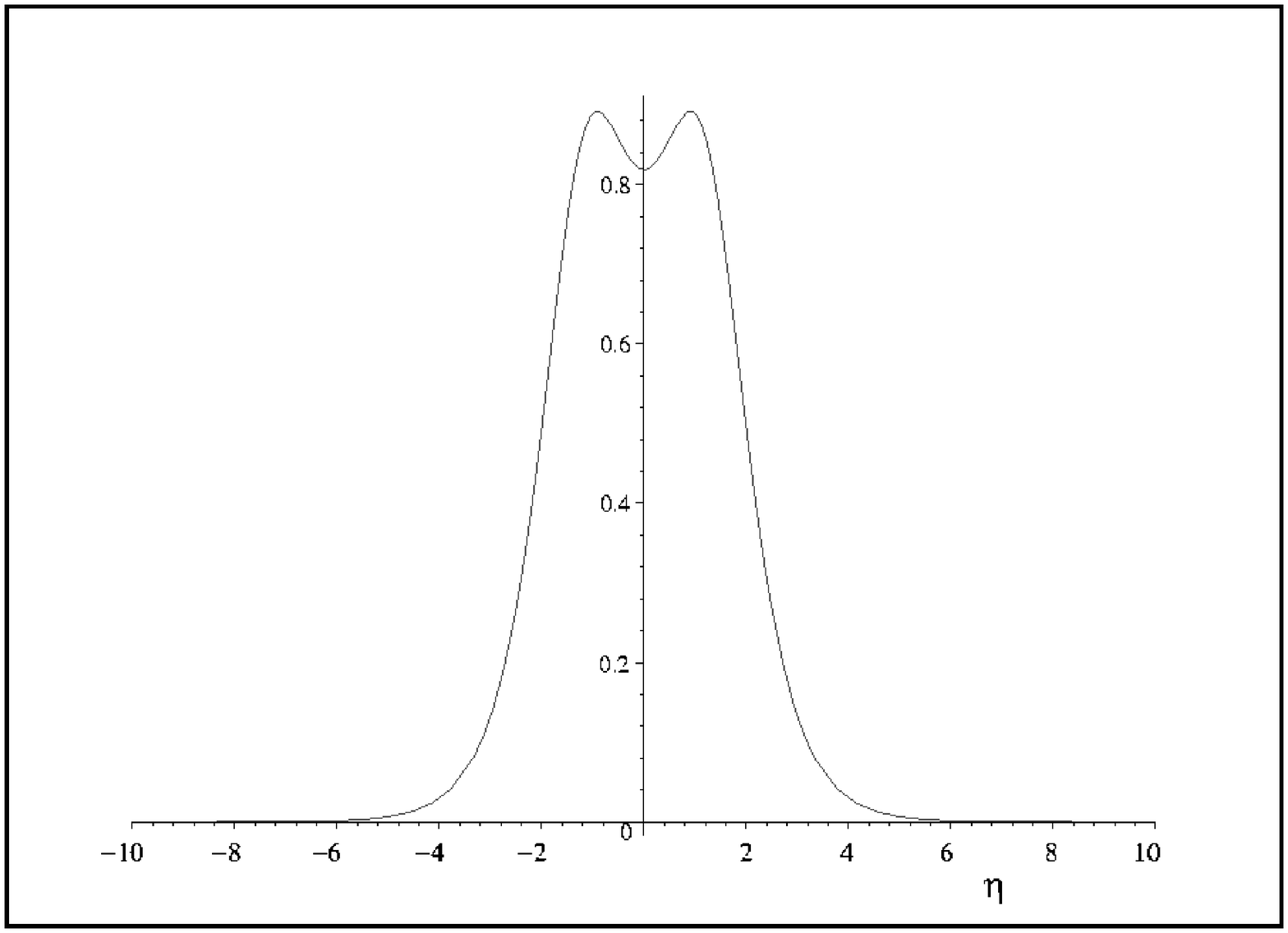}
\caption{Bulk dependence of the kinetic energy term,
Eq. \ref{dimensionless normalisation integral}, for left-handed
fermions on a domain wall with AdS$_4$ brane cosmology;
$A=0.2,B=0.8,h_F=0.1$; see Eqs. \ref{locally localised gravity ansatz}-\ref{locally localised V} 
and Eq. \ref{h_F and mu}.}
\label{leftfermionunconfined}}

\section{Conclusion}

\label{conclusion}

We have presented solutions to the Einstein equations describing a
scalar field coupled to five-dimensional gravity, with a warped
five-dimensional metric and dS$_4$ and AdS$_4$ brane cosmology. In the
dS$_4$ case, the metric necessarily contains zeroes which generally
(but not inevitably) correspond to curvature singularities. It is possible to obtain a warped metric
with dS$_4$ slices from a smooth potential and scalar field, and in
this case the metric zeroes are simply horizons. However, there are
stringent conditions on the metric in this case which make it
difficult to write down an analytic solution. Moreover, the presence of zeroes
in the five-dimensional warped metric, whether representing horizons or curvature
singularities, generally leads to divergences in the normalisation
integrals for both the left- and right-handed zero modes of the Dirac
field. This issue does not arise in the case of an infinitely thin
brane, where the matter fields are confined a priori to a
3+1-dimensional slice of the higher-dimensional space and do not
extend into the bulk at all.

In the AdS$_4$ case, we have derived new  analytic
self-consistent solutions for the metric warp factor and scalar field kink, motivated by the principle
that localisation of gravity should depend only on the behaviour of
the metric close to the brane (\cite{karch+randall}). Although for non-trivial
brane cosmology the scalar field does not generally seem to asymptote
to minima of the potential, these configurations (and their dS$_4$
counterparts) are expected to be both classically stable and to confine
the four-dimensional graviton \cite{soda}.
The ensuing
warped metric admits both left- and right-handed normalisable fermionic
zero modes (as in  \cite{koley+kar2, bajc+gabadadze}), although not all the normalisable modes are localised on
the brane. At present we have only investigated massless chiral
fermions, but it might also be interesting to investigate the spectrum of massive
fermionic modes in these backgrounds, using Eqs. \ref{massive
modes 1}-\ref{massive modes 2}.

Our present analysis only applies to a particular class of warped
metrics, corresponding to four-dimensional metrics of constant
curvature. Future studies dealing with the effects of matter on the
brane will need to employ a more general metric ansatz, allowing
metric elements that are non-separable functions of
the brane and bulk coordinates.

\acknowledgments

This work was supported in part by the Australian Research
Council and in part by the University of Melbourne through a Grimwade Scholarship.

\end{document}